# Optical monitoring in southern blazars. Analysis of variability and spectral colour behaviours

L. Zibecchi ,[1,2]★† I. Andruchow,[1,3] E. J. Marchesini ,[4] S. A. Cellone[1,5] and J. A. Combi[1,3,6]

[1]*Facultad de Ciencias Astronómicas y Geofísicas, Universidad Nacional de La Plata, Paseo del Bosque, B1900FWA La Plata, Provincia de Buenos Aires, Argentina*
[2]*Instituto de Astrofísica de La Plata (IALP), CONICET-UNLP, B1900FWA La Plata, Provincia de Buenos Aires, Argentina*
[3]*Instituto Argentino de Radioastronomía, CONICET-CICPBA-UNLP, CC5 (1894) Villa Elisa, Provincia de Buenos Aires, Argentina*
[4]*INAF–Osservatorio di Astrofisica e Scienza dello Spazio, Via Gobetti 93/3, I-40129 Bologna, Italy*
[5]*Complejo Astronómico El Leoncito (CASLEO), CONICET-UNLP-UNC-UNSJ, J5402DSP, Provincia de San Juan, Argentina*
[6]*Departamento de Física (EPS), Universidad de Jaén, Campus Las Lagunillas s/n, A3, E-23071 Jaén, Spain*

**ABSTRACT**
We present the results of the optical monitoring of 18 southern blazars in the *V* and *R* Johnson–Cousins bands. Our main objective is to study the variations in the optical flux and their relationship with the colour and spectral variabilities. The optical observations were acquired with the 2.15 m 'Jorge Sahade' telescope, CASLEO, Argentina. The whole campaign comprised from 2014 April to 2019 September. In addition, X-ray data were taken from the *Chandra* X-ray Observatory and the *Swift*/XRT databases, and $\gamma$-ray data were taken from the *Fermi*-Large Area Telescope 3FGL catalogue. From the total of 18 blazars, we found variability in each one of the time-scales considered for 6 blazars (PKS 0208–512, PKS 1116–46, PKS 1440–389, PKS 1510–089, PKS 2005–489, and PKS 2155–304). In particular, from the colour–magnitude and the multiwavelength analysis, we found that PKS 1510-089 (flat-spectrum radio quasar) is undergoing an activity phase. For the case of PKS 2005–489 (BL Lac), this blazar is in a quiescent state, in which it has been for more than a decade, and it is compatible with its *bluer-when-brighter* moderate tendency, possibly due to the presence of shocks within the jet.

## 1 INTRODUCTION

Among all the different sub-classes of active galactic nuclei (AGNs), blazars are the most extreme one, having extremely collimated relativistic jets pointing very close (<10°) to the line of sight of the observer (Urry & Padovani 1995). This implies relativistic boosting of the jet emission, dominating the blazar emission, which spans the entire electromagnetic spectrum (from radio to $\gamma$-rays). The spectral energy distributions (SEDs) are bi-modal with two peaks: the first one between sub-millimeter and UV wavelengths, and the other at high energies (MeV–GeV) corresponding to the X/$\gamma$-rays region of the spectrum. The blazar emission can be well described by the *synchrotron self-Compton* (SSC) models, i.e. the jet accelerates electrons that emit light via synchrotron process – responsible of the low-energy peak – and these photons are scattered to higher energies via the inverse Compton process with the same accelerated electrons – the high-energy peak – Tavecchio, Maraschi & Ghisellini (1998) Böttcher (2007), Dermer et al. (2009), and leptonic models. The high-energy emission in blazars can be alternatively explained by hadronic processes, or by the presence of an external Compton component, i.e. another source of photons to be scattered to higher energies, other than those of the jet itself. Since the publication of the first *Fermi*-LAT catalogue (*Large Area Telescope,* Abdo et al. 2010c), the blazars, classified into BL Lacs and flat-spectrum radio quasars (FSRQs), are sub-divided into three SED classes, based on the frequency at which the synchrotron peak is located: *low synchrotron peak* (LSP) blazars, *intermediate synchrotron peak* (ISP) blazars, and *high synchrotron peak* (HSP) blazars.

Blazars are also characterized by their intense and variable emission across the electromagnetic spectrum, from radio to $\gamma$-ray wavelengths. The variability observed in blazars is one of their defining characteristics and could be related to processes occurring in their relativistic jets. These variations are detected at different time-scales. Blazars can exhibit rapid variability on time-scales as short as hours, known as *intraday variability* (IDV). This rapid variability challenges models of emission mechanisms and suggests compact regions within the jets undergoing rapid changes. IDV has been extensively observed in various blazars across different wavelengths, including optical, X- and $\gamma$-rays bands (Wagner & Witzel 1995; Marscher et al. 2002; Gupta & Joshi 2005; Abdo et al. 2010b). This type of variability (also known as microvariability) allows to estimate the size of the emission region and to conjecture the possible value of the mass of the central super massive black hole hosted in these objects (Gupta, Srivastava & Wiita 2009; Liu &

★ E-mail: lorenazibecchi@gmail.com
† Visiting Astronomer, Complejo Astronómico El Leoncito operated under agreement between the Consejo Nacional de Investigaciones Científicas y Técnicas de la República Argentina and the National Universities of La Plata, Córdoba and San Juan.

Bai [2015]). This kind of variations can be detected from 1 GHz up to GeV energies, implying that an important part of the energy is emitted in regions smaller than 200 au, within the relativistic jet (Wagner & Witzel [1995]). Blazars also show variability on longer time-scales (Sandrinelli, Covino & Treves [2014]; Sandrinelli et al. [2018]), ranging from days to weeks (*short-term variability*, STV), and from months to years (*long-term variability*, LTV). In these cases, long-term monitoring campaigns have revealed flux variations and spectral changes in many blazars (Raiteri et al. [2005], [2007], [2009], [2021]). The study of LTV provides insights into the long-term behaviour of AGNs and their jet dynamics (Wagner & Witzel [1995]; Villata & Raiteri [1999]; Lähteenmäki et al. [2018]).

The analysis of the correlations between the colour index and the magnitude is a useful tool to study the nature of the microvariations present in the optical flux, since flux variations develop in colour changes and thus in spectral changes (Gu & Ai [2011]; Agarwal et al. [2016]; Xiong et al. [2016]). From individual studies of different sources, it was found that blazars such as OJ 287, BL Lacertae, 3C 273, PG 1553+113, among others (Takalo & Sillanpaa [1989]; Ghisellini et al. [1997]; Villata et al. [2002], [2004]; Papadakis, Villata & Raiteri [2007]; Dai et al. [2009], [2011]; Raiteri et al. [2015]; Gupta et al. [2016]; Man et al. [2016]; Li et al. [2017]; Weaver et al. [2020]) exhibited a *bluer-when-brighter* (BWB) trend, meanwhile other blazars showed a *redder-when-brighter* (RWB) trend, such as PKS 0736+017, 3C 454.3, and PKS 0208–512 (Ramírez et al. [2004]; Villata et al. [2006]; Chatterjee, Nalewajko & Myers [2013b]; Sarkar et al. [2019a]). Additionally, studies on large samples of blazars have indicated that the BWB tendency is commonly found in BL Lacs, while FSRQs show an RWB trend (Vagnetti, Trevese & Nesci [2003]; Gu et al. [2006a]; Rani et al. [2010]; Ikejiri et al. [2011]; Bonning et al. [2012]; Wierzcholska et al. [2015]). The physical mechanisms behind these trends are related to the emission processes occurring within the jets and the accretion disc around the central supermassive black hole. The RWB trend is typically associated with a fixed bluer disc component and a variable redder jet one. The bluer disc component is assumed to be relatively stable over time, while the variability in the redder jet component is what drives the RWB trend. This variability in the jet emission could arise due to changes in the relativistic beaming effect caused by the motion of plasma within the jet or changes in the physical conditions (e.g. magnetic field strength and particle density) within the jet itself. Conversely, the BWB trend is characterized by bluer colours when the source is brighter. This tendency could be explained by variations in the high-energy emission processes within the jet dominating the overall SED. As the jet becomes brighter, it could lead to an increase in the high-energy emission, which tends to be bluer in spectral colour (Jorstad et al. [2001]; Raiteri et al. [2007]; Pacciani et al. [2014]).

Other observational studies revealed more complicated correlations between magnitudes and colours. For example, five different kinds of behaviours were found in the colour–magnitude diagrams according to Zhang et al. ([2014]): BWB, RWB, and RWB behaviour when the blazar has its lowest state of activity, BWB when the state of activity is the highest (RWB to BWB); stable when brighter (SWB); and no correlation with the brightness of the source. Recently, Zhang, Zhao & Wu ([2022]) found two new spectral behaviours: the *redder-stable-when-brighter* and the *bluer-stable-when-brighter* trends. The first one corresponds to a blazar which shows an RWB tendency in its low state and a stability in its high state. In this case, the spectrum in the optical band becomes steeper when the blazars brightens in the low state and keeps stable (without variations) in the high state. On the contrary, the second new trend consists in a blazar exhibiting a BWB trend in the low state, while keeping invariable in the high state. Observing the whole panorama, a rigorous framework of flux–colour correlation patterns in blazars has not yet been established, and, thus, more multicolour observations are needed to deeply study these correlations.

As mentioned before, blazars show extreme variations in other wavelengths. In particular, variability in X- and $\gamma$-ray bands is a prominent feature that provides valuable insights into the physical processes occurring in their relativistic jets. X-ray variability in blazars is often rapid and can occur on time-scales ranging from minutes to days, and sometimes even shorter. This rapid variability suggests that the emission region responsible for X-rays is compact and close to the supermassive black hole at the centre of the AGN. These variations are associated with changes in the number, energy distribution, and acceleration mechanisms of relativistic electrons within the jet. These changes can be triggered by shocks, magnetic reconnection events, or instabilities in the jet's magnetic field (Böttcher, Baring & Summerlin [2012]; Böttcher et al. [2013]). On the other hand, $\gamma$-ray variability is observed across a wide range of time-scales, from minutes to years. These variations can result from changes in the population of relativistic electrons, variations in the seed photon population (e.g. synchrotron photons), or changes in the Doppler boosting factor due to variations in the jet orientation or velocity (H. E. S. S. Collaboration [2011], [2017], [2021]). Another remarkable aspect of blazars is that they are the most common very high-energy (VHE; $E > 100$ GeV) $\gamma$-ray sources in the Universe. Of the more than 250 sources detected in the VHE band to date with the Imaging Atmospheric Cherenkov Telescopes,[1] about 35 per cent are extragalactic sources (particularly, AGNs) and 95 per cent of them are classified as blazars. In this energy band, blazars are extremely variable (H. E. S. S. Collaboration [2010], [2013]). Thus, studying correlations between variabilities in different frequency in blazars is crucial for understanding their emission mechanisms and the physical processes occurring in their jets. Observations have revealed simultaneous variability across multiple wavelengths in blazars, suggesting a connection between emissions in different bands. For instance, variations seen in optical fluxes often correspond to changes in X-ray and gamma-ray emissions within the same period (Singh et al. [2019]; Kapanadze [2021]; Khatoon et al. [2022]; Tolamatti et al. [2022]; Yuan et al. [2023]). These studies can provide clues on the acceleration and emission processes of relativistic particles, which are thought to occur in a region near to the central super massive black hole (Neronov & Aharonian [2007]; Rieger & Aharonian [2008]; Istomin & Sol [2009]) and/or also in the jet (Katarzyński, Sol & Kus [2001]; Mücke & Protheroe [2001]; Katarzyński, Sol & Kus [2003]; Böttcher [2007]; Tavecchio & Ghisellini [2008]).

In the present paper, we analysed the light curves and the colour behaviour of a sample of 18 high-energy-detected blazars in the optical band. In Section [2], we describe the blazar sample and the procedure of the data reduction. In Section [3], we show the analysis tools and in Section [4] we present the results and their analysis and discussion. In Section [5], we present our conclusions.

## 2 OBSERVATIONS AND DATA REDUCTION

We selected a sample of 18 blazars from the 5th edition of the Roma–BZCAT catalogue[2] (Massaro et al. [2014]). This sample is composed by BL Lacs, FSRQs, and BZUs (blazars of uncertain type). The latter ones are sources that exhibit blazar activity but also have peculiar characteristics, such as presence or absence of broad spectral lines or other features, transition objects between a radio galaxy and a BL Lac, or galaxies hosting a low luminosity

---

[1] http://tevcat.uchicago.edu/
[2] http://www.asdc.asi.it/bzcat/

**Table 1.** Blazar sample. $\gamma$-rays.

| Object | Obj. class | SED | $\alpha$ (J2000.0) h m s | $\delta$ (J2000.0) ° ′ arcsec | z | m mag | X-Ray | $\gamma$-Ray |
|---|---|---|---|---|---|---|---|---|
| PKS 0208−512 | BZU[a] | — | 02 10 46.2 | −51 01 01.8 | 1.003 | 14.82(R) | ✓[d] | ✓ |
| [HB89] 0414+009 | BLLac | HSP | 04 16 52.5 | +01 05 23.9 | 0.287 | 15.86(R) | ✓[d] | ✗ |
| PKS 0521−36 | BZU[a] | — | 05 22 57.9 | −36 27 30.8 | 0.057 | 14.48(R) | ✓[d] | ✓ |
| 3FGL J0846.9−2336 | BZU[a] | — | 08 47 01.5 | −23 37 01.6 | 0.061 | 13.00(R) | ✓[c] | ✓ |
| PKS 1116−46 | FSRQ | LSP | 11 18 26.9 | −46 34 15.0 | 0.713 | 17.02(V) | ✗ | ✓ |
| PKS 1127−14 | FSRQ | LSP | 11 30 07.1 | −14 49 27.4 | 1.188 | 16.70(V) | ✓[d] | ✓ |
| PKS 1229−02 | FSRQ | LSP | 12 32 00.0 | −02 24 05.3 | 1.043 | 16.80(R)[b] | ✓[d] | ✓ |
| PMN J1256−1146 | BLLac | HSP | 12 56 15.9 | −11 46 37.4 | 0.058 | 11.01(R) | ✓[c] | ✓ |
| PKS 1424−41 | FSRQ | LSP | 14 27 56.3 | −42 06 19.4 | 1.522 | 16.30(R) | ✓[c] | ✓ |
| PKS 1440−389 | BLLac | HSP | 14 43 57.2 | −39 08 39.7 | 0.139 | 14.81(V) | ✗ | ✓ |
| PKS 1510−089 | FSRQ | LSP | 15 12 50.5 | −09 05 59.8 | 0.360 | 16.10(R) | ✓[d] | ✓ |
| 3FGL J1917.7−1921 | BLLac | HSP | 19 17 44.8 | −19 21 31.6 | 0.137 | 15.24(R) | ✓[c] | ✓ |
| 3FGL J1958.2−3011 | BLLac | HSP | 19 58 14.9 | −30 11 11.8 | 0.119 | 13.97(R) | ✓[d] | ✓ |
| PKS 2005−489 | BLLac | HSP | 20 09 25.3 | −48 49 53.7 | 0.071 | 11.41(R) | ✓[d] | ✓ |
| PKS 2126−158 | FSRQ | LSP | 21 29 12.1 | −15 38 41.0 | 3.268 | 16.43(R) | ✓[d] | ✗ |
| PKS 2149−306 | FSRQ | LSP | 21 51 55.5 | −30 27 53.6 | 2.340 | 17.48(R) | ✓[d] | ✓ |
| PKS 2155−304 | BLLac | HSP | 21 58 52.0 | −30 13 32.1 | 0.117 | 12.62(R) | ✓[d] | ✓ |
| PMN J2310−4374 | BLLac | – | 23 10 41.7 | −43 47 34.1 | 0.088 | 15.92(V) | ✓[d] | ✗ |

[a]Uncertain type blazar. [b]From the Roma-BZCAT, 5th edition. [c]Observed with *Swift*. [d]Observed with *Chandra* and *Swift*. *Note.* In column 1, we list the object name, in Column 2 object class, in column 3 SED class, in columns 4 and 5 right ascension and declination, in column 6 redshift, in column 7 the optical band magnitudes taken from NED, and in columns 8 and 9 the data available for X- and -rays.

**Table 2.** Variability index values given by the GLVARY tool.

| Variability index | Condition | Result |
|---|---|---|
| 0 | $P \leq 0.5$ | Definitely not variable |
| 1 | $0.5 < P < 2/3$ AND $f3 > 0.997$ AND $f5 = 1.0$ | Considered not variable |
| 2 | $2/3 \leq 0.9$ AND $f3 > 0.997$ AND $f5 = 1.0$ | Probably not variable |
| 3 | $0.5 \leq P < 0.6$ | May be variable |
| 4 | $0.6 \leq P < 2/3$ | Likely to be variable |
| 5 | $2/3 \leq P < 0.9$ | Considered variable |
| 6 | $0.9 \leq P$ AND Odd $< 2.0$ | Definitely variable |
| 7 | $2.0 \leq$ Odd $< 4.0$ | Definitely variable |
| 8 | $4.0 \leq$ Odd $< 10.0$ | Definitely variable |
| 9 | $10.0 \leq$ Odd $< 30.0$ | Definitely variable |
| 10 | $30.0 \leq$ Odd | Definitely variable |

*Note.* Table taken from http://cxc.cfa.harvard.edu/ciao/threads/variable/.

blazar nucleus (Massaro et al. 2009). In particular, we looked for those sources whose characteristics in the optical band (position on the sky, magnitude, etc.) allowed their follow-up from CASLEO, Argentina, in the Southern Hemisphere. If possible, these objects should have observations made with the gamma-ray satellite *Fermi*-LAT and, in a complementary way, with an X-ray satellite, such as *Chandra* or *Swift*. On the other hand, we were interested in building a sample including both known and well-studied sources, as well as sources that have not been studied, especially in terms of variability analysis and classification. In Table 1, we show the properties of the sample: the name of the source, the object and SED classes, its coordinates, the redshift, the magnitude (extracted from NED[3] – *NASA/IPAC Extragalactic Database*) and the information about the X- and $\gamma$- rays data available on public data bases.

### 2.1 Optical data

The optical images were taken with the 2.15 m 'Jorge Sahade' telescope, at CASLEO, Argentina. The whole campaign comprised from 2014 April to 2019 September, divided into seven observational periods. The telescope is equipped with a CCD camera, using a Roper chip with a gain of 2.18 electrons/adu and a *read-out-noise* of 3.5 electrons. This detector has an area of $2048 \times 2048$ pixels of 13.5 $\mu$m each side, installed in a Dewar cooled with liquid nitrogen. Most of the observations were made using a focal-reducer, resulting in a scale of 0.45 arcsec per pixel, which provides a field of view (FoV) of $9 \times 9$ arcmin$^2$. The exception is the 2014A run, in which we did not use the focal-reducer, obtaining an FoV of $5.2 \times 5.2$ arcmin$^2$, with a scale of 0.15 arcsec per pixel. We obtained photometric images in the $V$ and $R$ Johnson–Cousins filters system for all the sources and the exposure times spanned from 90 to 600 s, depending on the brightness of the object as well as weather and atmospheric conditions. We used the standard procedures within the IRAF[4] reduction package to obtain the images corrected by bias and flat-fields. For the photometry, we used the IRAF APPHOT package with different aperture radii ($r_{ap}$)

---

[3]https://ned.ipac.caltech.edu/

[4]IRAF is distributed by the National Optical Astronomy Observatories, which are operated by the Association of Universities for Research in Astronomy, Inc., under cooperative agreement with the National Science Foundation.

Table 3. Results for the variability state of the sources in the optical band.

| Object | Date mm/dd/yyyy | C V | C R | F-area V | F-area R | Var. | N | Γ V | Γ R | $\sigma_2$ V | $\sigma_2$ R | BWB or RWB | ⟨V⟩ mag | ⟨R⟩ mag | $\Delta m_V$ | $\Delta m_R$ | $\alpha_{VR}$ |
|---|---|---|---|---|---|---|---|---|---|---|---|---|---|---|---|---|---|
| PKS 0208−512 | 08/13/2015 | 1.983 | 4.372 | 0.930 | 0.999 | Yes$_R$ | 9 | 1.53 | 1.54 | 0.016 | 0.008 | BWB (0.287) | 16.95 (0.01) | 16.58 (0.01) | 0.102 | 0.098 | 2.11 (0.01) |
| | 09/15/2015 | 1.087 | 0.673 | 0.141 | 0.599 | No | 6 | 2.34 | 2.70 | 0.052 | 0.030 | | 17.47 (0.01) | 17.20 (0.01) | 0.141 | 0.060 | 1.54 (0.01) |
| | 09/17/2015 | 0.580 | 0.683 | 0.826 | 0.664 | No | 8 | 2.39 | 2.67 | 0.030 | 0.017 | | 17.52 (0.01) | 17.26 (0.01) | 0.059 | 0.037 | 1.48 (0.01) |
| | All nights Sept. | 0.828 | 0.931 | 0.493 | 0.201 | No | 14 | 2.35 | 2.71 | 0.053 | 0.036 | | 17.49 (0.01) | 17.23 (0.01) | 0.141 | 0.103 | 1.51 (0.01) |
| | All nights | 5.872 | 11.48 | 1.000 | 1.000 | Yes | 23 | 2.20 | 2.51 | 0.047 | 0.028 | RWB (−0.759) | 17.28 (0.01) | 16.98 (0.01) | 0.659 | 0.743 | 1.81 (0.01) |
| [HB89] 0414+009 | 11/27/2016 | 1.744 | 1.575 | 0.935 | 0.871 | No | 13 | 0.59 | 0.70 | 0.009 | 0.006 | | 16.19 (0.01) | 15.77 (0.01) | 0.058 | 0.033 | 2.37 (0.02) |
| | 11/28/2016 | 2.011 | 2.057 | 0.971 | 0.968 | No | 11 | 0.60 | 0.70 | 0.007 | 0.005 | | 16.20 (0.01) | 15.77 (0.01) | 0.048 | 0.034 | 2.43 (0.02) |
| | All nights Nov. | 1.953 | 1.579 | 0.998 | 0.967 | No | 24 | 0.59 | 0.70 | 0.008 | 0.006 | | 16.20 (0.01) | 15.77 (0.01) | 0.074 | 0.034 | 2.40 (0.02) |
| PKS 0521−36 | 12/11/2015 | 1.785 | 0.838 | 0.987 | 0.562 | No | 21 | 0.99 | 0.96 | 0.008 | 0.015 | | 16.80 (0.01) | 16.47 (0.01) | 0.054 | 0.059 | 1.85 (0.02) |
| 3FGL J0846.9−2336 | 04/22/2014 | 1.476 | 1.728 | 0.788 | 0.917 | No | 12 | 0.92 | 0.75 | 0.008 | 0.007 | | 16.69 (0.01) | 16.06 (0.01) | 0.041 | 0.035 | 3.56 (0.01) |
| | 04/23/2014 | 2.170 | 0.645 | 0.650 | 0.412 | No | 3 | 0.92 | 0.75 | 0.001 | 0.002 | | 16.67 (0.01) | 16.07 (0.01) | 0.010 | 0.007 | 3.45 (0.01) |
| | 04/24/2014 | 1.046 | 1.341 | 0.067 | 0.416 | No | 5 | 0.93 | 0.74 | 0.013 | 0.005 | | 16.66 (0.01) | 16.05 (0.01) | 0.013 | 0.015 | 3.46 (0.01) |
| | 04/25/2014 | 0.712 | 0.878 | 0.472 | 0.192 | No | 5 | 0.91 | 0.73 | 0.013 | 0.010 | | 16.67 (0.01) | 16.06 (0.01) | 0.031 | 0.014 | 3.46 (0.01) |
| | 04/26/2014 | 1.968 | 0.298 | 0.589 | 0.837 | No | 4 | 0.95 | 0.74 | 0.010 | 0.013 | | 16.69 (0.02) | 16.07 (0.02) | 0.031 | 0.007 | 3.51 (0.01) |
| | 04/27/2014 | 0.724 | 1.110 | 0.549 | 0.194 | No | 7 | 0.96 | 0.77 | 0.007 | 0.007 | | 16.71 (0.01) | 16.09 (0.01) | 0.023 | 0.025 | 3.51 (0.01) |
| | All nights April | 1.700 | 2.768 | 0.997 | 1.000 | No | 36 | 0.92 | 0.74 | 0.016 | 0.009 | | 16.68 (0.01) | 16.07 (0.01) | 0.067 | 0.068 | 3.52 (0.01) |
| PKS 1116−46 | 04/13/2015 | – | 1.294 | – | 0.653 | No | 15 | – | 1.16 | – | 0.010 | | – | 16.93 (0.01) | – | 0.050 | – |
| | 04/14/2015 | 4.322 | 2.729 | 1.000 | 0.998 | Yes | 12 | 0.90 | 1.12 | 0.009 | 0.010 | BWB (0.540) | 16.98 (0.01) | 16.88 (0.01) | 0.103 | 0.063 | 0.53 (0.01) |
| | All nights | 4.322 | 3.374 | 1.000 | 1.000 | Yes | 27 | 0.90 | 1.14 | 0.009 | 0.010 | BWB (0.541) | 16.98 (0.01) | 16.90 (0.01) | 0.103 | 0.139 | 0.53 (0.01) |
| PKS 1127−14 | 04/10/2015 | 1.218 | 1.014 | 0.531 | 0.418 | No | 15 | 1.09 | 1.19 | 0.005 | 0.006 | | 17.72 (0.01) | 17.27 (0.01) | 0.023 | 0.027 | 2.50 (0.01) |
| | 04/12/2015 | 1.109 | 0.992 | 0.225 | 0.174 | No | 9 | 1.08 | 1.19 | 0.005 | 0.005 | | 17.72 (0.01) | 17.28 (0.01) | 0.017 | 0.017 | 2.44 (0.01) |
| | All nights Apr. | 0.884 | 0.957 | 0.438 | 0.165 | No | 24 | 1.09 | 1.18 | 0.007 | 0.006 | | 17.71 (0.01) | 17.27 (0.01) | 0.023 | 0.027 | 2.49 (0.01) |
| PKS 1229−02 | 04/12/2015 | 0.669 | 0.807 | 0.820 | 0.531 | No | 13 | 0.74 | 1.05 | 0.009 | 0.007 | | 16.79 (0.01) | 16.48 (0.01) | 0.021 | 0.017 | 1.76 (0.01) |
| PMN J1256−1146 | 04/23/2014 | 0.512 | 1.239 | 0.872 | 0.385 | No | 7 | 0.41 | 0.35 | 0.003 | 0.002 | | 15.64 (0.02) | 15.06 (0.02) | 0.010 | 0.015 | 3.29 (0.02) |
| PKS 1424−41 | 04/14/2015 | 2.533 | 1.575 | 0.983 | 0.780 | No | 9 | 0.99 | 0.94 | 0.016 | 0.010 | | 17.63 (0.01) | 17.05 (0.01) | 0.111 | 0.040 | 3.30 (0.01) |
| PKS 1440−389 | 04/24/2014 | 1.329 | 1.335 | 0.932 | 0.935 | No | 43 | 0.83 | 0.85 | 0.004 | 0.005 | | 15.08 (0.01) | 14.66 (0.01) | 0.020 | 0.025 | 2.37 (0.01) |
| | 04/25/2014 | 2.014 | 2.297 | 0.998 | 0.999 | No | 25 | 0.84 | 0.87 | 0.007 | 0.006 | | 15.10 (0.01) | 14.70 (0.01) | 0.033 | 0.038 | 2.29 (0.01) |
| | 04/27/2014 | 3.439 | 2.097 | 0.999 | 0.979 | Yes$_V$ | 12 | 0.82 | 0.84 | 0.007 | 0.009 | BWB (0.870) | 15.08 (0.01) | 14.67 (0.01) | 0.060 | 0.042 | 2.34 (0.01) |
| | All nights | 2.833 | 2.360 | 1.000 | 1.000 | Yes | 80 | 0.83 | 0.85 | 0.006 | 0.008 | RWB (−0.176) | 15.08 (0.02) | 14.67(0.02) | 0.066 | 0.066 | 2.34 (0.01) |
| PKS 1510−089 | 04/12/2015 | 0.914 | 2.176 | 0.259 | 0.993 | No | 15 | 0.98 | 0.95 | 0.007 | 0.005 | | 15.41 (0.01) | 14.91 (0.01) | 0.021 | 0.048 | 2.81 (0.01) |
| | 04/13/2015 | – | 1.757 | – | 0.938 | No | 13 | – | 0.96 | – | 0.009 | | – | 14.95 (0.01) | – | 0.046 | – |
| | 04/14/2015 | 2.571 | 2.746 | 0.996 | 0.997 | Yes | 12 | 0.99 | 0.95 | 0.007 | 0.007 | BWB (0.018) | 15.42 (0.01) | 14.93 (0.01) | 0.049 | 0.054 | 2.75 (0.01) |
| | All nights 2015 | 1.876 | 2.839 | 0.998 | 1.000 | Yes | 40 | 0.99 | 0.96 | 0.007 | 0.007 | RWB (−0.146) | 15.41 (0.01) | 14.92 (0.01) | 0.049 | 0.070 | 2.78 (0.01) |
| | 04/04/2019 | 0.335 | 1.582 | 0.982 | 0.711 | No | 7 | 1.33 | 1.35 | 0.019 | 0.007 | | 15.88 (0.01) | 15.48 (0.01) | 0.021 | 0.036 | 2.30 (0.01) |
| | 04/05/2019 | 2.053 | 2.779 | 0.989 | 0.999 | Yes$_R$ | 15 | 1.44 | 1.47 | 0.009 | 0.006 | BWB (0.483) | 15.98 (0.01) | 15.59 (0.01) | 0.091 | 0.080 | 2.21 (0.01) |
| | 04/06/2019 | 1.397 | 0.816 | 0.603 | 0.395 | No | 8 | 1.33 | 1.34 | 0.003 | 0.004 | | 15.89 (0.01) | 15.49 (0.01) | 0.013 | 0.012 | 2.28 (0.01) |
| | 04/07/2019 | 1.930 | 3.694 | 0.961 | 0.999 | Yes$_R$ | 12 | 1.20 | 1.19 | 0.006 | 0.003 | BWB (0.683) | 15.74 (0.01) | 15.31 (0.01) | 0.042 | 0.039 | 2.47 (0.01) |
| | All nights Apr. | 6.906 | 14.095 | 1.000 | 1.000 | Yes | 42 | 1.41 | 1.43 | 0.010 | 0.006 | RWB (−0.850) | 15.88 (0.01) | 15.47 (0.01) | 0.304 | 0.336 | 2.25 (0.01) |
| | 05/07/2019 | 1.805 | 3.104 | 0.976 | 1.000 | Yes$_R$ | 17 | 0.92 | 0.92 | 0.005 | 0.004 | BWB (0.023) | 15.29 (0.01) | 14.87 (0.01) | 0.030 | 0.037 | 2.39 (0.01) |
| | 05/08/2019 | 1.209 | 1.062 | 0.514 | 0.175 | No | 15 | 0.83 | 0.83 | 0.006 | 0.006 | | 15.09 (0.01) | 14.65 (0.01) | 0.021 | 0.020 | 2.47 (0.01) |
| | 05/09/2019 | 0.778 | 0.893 | 0.603 | 0.299 | No | 13 | 0.89 | 0.90 | 0.008 | 0.004 | | 15.24 (0.01) | 14.83 (0.01) | 0.021 | 0.021 | 2.35 (0.01) |
| | 05/10/2019 | 0.566 | 1.218 | 0.940 | 0.515 | No | 13 | 0.97 | 0.98 | 0.007 | 0.004 | | 15.40 (0.01) | 14.98 (0.01) | 0.017 | 0.017 | 2.37 (0.01) |
| | All nights May | 18.41 | 25.26 | 1.000 | 1.000 | Yes | 58 | 0.91 | 0.92 | 0.007 | 0.005 | RWB (−0.631) | 15.25 (0.01) | 14.83 (0.01) | 0.330 | 0.349 | 2.40 (0.01) |
| | All nights 2019 | 41.27 | 61.07 | 1.000 | 1.000 | Yes | 100 | 0.98 | 0.99 | 0.008 | 0.006 | RWB (−0.616) | 15.52 (0.03) | 15.09 (0.04) | 0.953 | 0.989 | 2.36 (0.02) |

Table 3 – *continued*

| Object | Date mm/dd/yyyy | C V | C R | F-area V | F-area R | Var. | N | Γ V | Γ R | $\sigma_2$ V | $\sigma_2$ R | BWB or RWB | $\langle V \rangle$ mag | $\langle R \rangle$ mag | $\Delta m_V$ | $\Delta m_R$ | $\alpha_{VR}$ |
|---|---|---|---|---|---|---|---|---|---|---|---|---|---|---|---|---|---|
| 3FGL J1917.7−1921 | Total | 23.26 | 21.77 | 1.000 | 1.000 | Yes | 140 | 1.11 | 1.24 | 0.011 | 0.011 | RWB (−0.400) | 15.49 (0.03) | 15.05 (0.03) | 0.953 | 0.989 | 2.48 (0.02) |
| | 04/25/2014 | 0.703 | 1.108 | 0.764 | 0.273 | No | 13 | 1.02 | 1.00 | 0.006 | 0.005 | | 15.53 (0.01) | 15.08 (0.01) | 0.014 | 0.022 | 2.55 (0.01) |
| | 04/26/2014 | 1.037 | 0.982 | 0.085 | 0.046 | No | 11 | 1.00 | 0.98 | 0.019 | 0.013 | | 15.56 (0.01) | 15.12 (0.01) | 0.060 | 0.047 | 2.51 (0.01) |
| | All nights Apr. | 1.642 | 2.422 | 0.976 | 0.999 | No | 24 | 1.01 | 0.98 | 0.014 | 0.009 | | 15.54 (0.01) | 15.10 (0.01) | 0.070 | 0.075 | 2.53 (0.01) |
| 3FGL J1958.2−3011 | 04/23/2014 | 2.382 | 1.008 | 0.989 | 0.020 | No | 11 | 0.88 | 0.75 | 0.004 | 0.009 | | 17.02 (0.01) | 16.44 (0.01) | 0.029 | 0.029 | 3.30 (0.01) |
| | 04/24/2014 | 0.861 | 1.344 | 0.336 | 0.609 | No | 10 | 0.87 | 0.75 | 0.004 | 0.007 | | 17.03 (0.02) | 16.44 (0.02) | 0.017 | 0.019 | 3.35 (0.02) |
| | All nights Apr. | 1.697 | 1.034 | 0.978 | 0.118 | No | 21 | 0.88 | 0.78 | 0.005 | 0.007 | | 17.02 (0.02) | 16.44 (0.02) | 0.034 | 0.029 | 3.32 (0.02) |
| PKS 2005−489 | 08/12/2015 | 1.601 | 1.444 | 0.972 | 0.922 | No | 26 | 0.84 | 0.80 | 0.006 | 0.005 | | 14.49 (0.04) | 14.08 (0.04) | 0.022 | 0.021 | 2.29 (0.04) |
| | 05/09/2019 | 0.423 | 1.182 | 0.945 | 0.305 | No | 7 | 0.78 | 0.74 | 0.007 | 0.004 | | 14.25 (0.04) | 13.83 (0.04) | 0.016 | 0.012 | 2.37 (0.04) |
| | 05/10/2019 | 2.863 | 1.136 | 0.987 | 0.255 | No | 8 | 0.77 | 0.73 | 0.003 | 0.004 | | 14.20 (0.04) | 13.82 (0.04) | 0.024 | 0.019 | 2.17 (0.04) |
| | All nights May | 2.606 | 1.821 | 0.999 | 0.967 | Yes$_V$ | 15 | 0.77 | 0.73 | 0.009 | 0.004 | BWB (0.982) | 14.22 (0.04) | 13.82 (0.04) | 0.068 | 0.030 | 2.26 (0.04) |
| | 08/23/2019 | 1.319 | 1.537 | 0.776 | 0.945 | No | 21 | 0.80 | 0.76 | 0.005 | 0.005 | | 14.36 (0.04) | 13.95 (0.04) | 0.024 | 0.021 | 2.32 (0.04) |
| | 08/24/2019 | 3.217 | 1.009 | 1.000 | 0.018 | Yes$_V$ | 26 | 0.81 | 0.77 | 0.004 | 0.005 | BWB (0.945) | 14.35 (0.04) | 13.96 (0.04) | 0.036 | 0.022 | 2.26 (0.04) |
| | 08/25/2019 | 1.673 | 1.878 | 0.983 | 0.997 | No | 24 | 0.81 | 0.77 | 0.004 | 0.004 | | 14.36 (0.04) | 13.97 (0.04) | 0.022 | 0.028 | 2.22 (0.04) |
| | 08/26/2019 | 1.878 | 1.769 | 0.997 | 0.994 | No | 26 | 0.81 | 0.77 | 0.005 | 0.004 | | 14.35 (0.04) | 13.96 (0.04) | 0.028 | 0.016 | 2.22 (0.04) |
| | 08/31/2019 | 1.58 | 1.56 | 0.958 | 0.952 | No | 22 | 0.82 | 0.78 | 0.004 | 0.004 | | 14.39 (0.04) | 13.98 (0.04) | 0.014 | 0.038 | 2.32 (0.04) |
| | 09/01/2019 | 1.331 | 1.780 | 0.801 | 0.989 | No | 21 | 0.81 | 0.78 | 0.004 | 0.004 | | 14.39 (0.04) | 13.98 (0.04) | 0.021 | 0.025 | 2.33 (0.04) |
| | 09/02/2019 | 1.323 | 1.405 | 0.803 | 0.881 | No | 23 | 0.81 | 0.77 | 0.006 | 0.004 | | 14.38 (0.04) | 13.97 (0.04) | 0.017 | 0.018 | 2.32 (0.04) |
| | 09/03/2019 | 0.599 | 0.546 | 0.983 | 0.994 | No | 24 | 0.80 | 0.77 | 0.009 | 0.008 | | 14.36 (0.04) | 13.95 (0.04) | 0.041 | 0.030 | 2.29 (0.04) |
| | All nights Aug.-Sept. | 2.013 | 2.148 | 1.000 | 1.000 | Yes | 187 | 0.81 | 0.78 | 0.009 | 0.009 | BWB (0.322) | 14.37 (0.04) | 13.96 (0.04) | 0.063 | 0.058 | 2.28 (0.04) |
| | All nights 2019 | 5.359 | 6.043 | 1.000 | 1.000 | Yes | 202 | 0.81 | 0.77 | 0.010 | 0.009 | | 14.35 (0.04) | 13.95 (0.04) | 0.210 | 0.186 | 2.28 (0.04) |
| | Total | 5.706 | 6.674 | 1.000 | 1.000 | Yes | 228 | 0.81 | 0.77 | 0.014 | 0.012 | BWB (0.287) | 14.37 (0.04) | 13.97 (0.04) | 0.307 | 0.283 | 2.28 (0.04) |
| PKS 2126−158 | 09/15/2015 | 1.634 | 1.051 | 0.912 | 0.141 | No | 14 | 1.02 | 1.10 | 0.010 | 0.016 | | 17.08 (0.03) | 16.81 (0.03) | 0.061 | 0.057 | 1.52 (0.03) |
| | 09/17/2015 | 0.361 | 0.513 | 0.998 | 0.963 | No | 12 | 0.99 | 1.10 | 0.017 | 0.015 | | 17.07 (0.02) | 16.82 (0.02) | 0.020 | 0.023 | 1.42 (0.02) |
| | All nights Sept. | 0.766 | 0.719 | 0.809 | 0.895 | No | 26 | 1.00 | 1.10 | 0.018 | 0.019 | | 17.07 (0.02) | 16.81 (0.02) | 0.061 | 0.057 | 1.48 (0.02) |
| PKS 2149−306 | 08/12/2015 | 1.208 | 0.427 | 0.236 | 0.692 | No | 4 | 1.58 | 1.85 | 0.028 | 0.103 | | 17.72 (0.03) | 17.53 (0.04) | 0.039 | 0.078 | 1.10 (0.05) |
| | 09/17/2015 | 1.025 | 0.785 | 0.037 | 0.349 | No | 5 | 1.54 | 1.79 | 0.013 | 0.014 | | 17.75 (0.01) | 17.54 (0.02) | 0.027 | 0.021 | 1.23 (0.03) |
| | All nights | 1.380 | 0.549 | 0.618 | 0.864 | No | 9 | 1.56 | 1.78 | 0.026 | 0.077 | | 17.74 (0.03) | 17.54 (0.04) | 0.058 | 0.099 | 1.18 (0.05) |
| PKS 2155−304 | 08/13/2015 | 2.983 | 2.940 | 1.000 | 1.000 | Yes | 19 | 0.82 | 0.83 | 0.010 | 0.012 | RWB (−0.741) | 12.95 (0.04) | 12.77 (0.04) | 0.077 | 0.099 | 1.03 (0.04) |
| | 09/15/2015 | 6.842 | 1.321 | 1.000 | 0.673 | Yes$_V$ | 14 | 0.60 | 0.62 | 0.007 | 0.012 | BWB (0.972) | 12.64 (0.04) | 12.47 (0.04) | 0.091 | 0.037 | 0.95 (0.03) |
| | All nights 2015 | 16.39 | 12.37 | 1.000 | 1.000 | Yes | 33 | 0.75 | 0.78 | 0.013 | 0.015 | BWB (0.406) | 12.82 (0.04) | 12.64 (0.04) | 0.420 | 0.370 | 1.01 (0.03) |
| PMN J2310−4374 | 09/17/2015 | 2.555 | 1.144 | 0.993 | 0.322 | No | 11 | 1.03 | 1.07 | 0.004 | 0.004 | | 16.38 (0.01) | 15.68 (0.01) | 0.039 | 0.015 | 3.94 (0.01) |

*Notes.* In column 1 we, list the object names, in column 2 the observation date, in columns 3–6 the *C* and *F* values in both filters, in column 7 the variability state, in column 8 the number of points *N*, in columns 9 and 10 the Γ factor in and , in columns 11 and 12 the dispersion $\sigma_2$, in column 13 the colour behaviour, in columns 14–17 the mean values of the and magnitudes and their variation amplitudes, and in column 18 the spectral index $\alpha_{RV}$. The errors for the and magnitudes as well as for the $\alpha_{RV}$ are given within the parenthesis

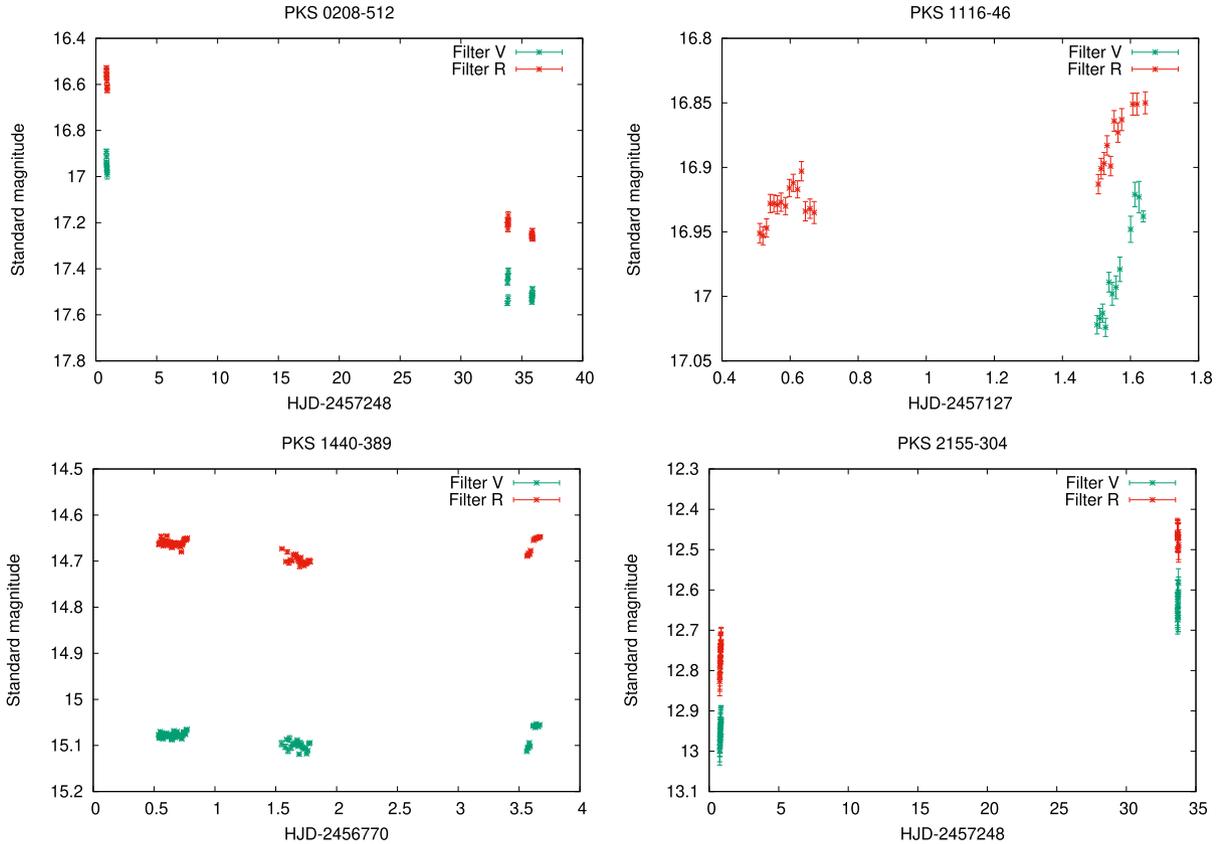

**Figure 1.** Standard magnitudes light curves in the *V* and *R* filters for PKS 0208−512, PKS 1116−46, PKS 1440−389, and PKS 2155−304.

depending on the source, its sky field and the stabilization of the photometric growth curve ($7 \leq r_{ap} \leq 18$ arcsec).

In order to obtain the standard *V* and *R* magnitudes for the blazars, we observed photometric standard star fields from Landolt's catalogue (Landolt 1992) in each observing run for each blazar follow-up. Additionally, we obtained the standard magnitudes for all the stars in the blazar fields (Zibecchi et al., in preparation).

### 2.2 X-ray data

For the different observations in the X-ray band, we used the *Chandra Data Archive*,[5] as well as the *Swift-XRT (Swift X-Ray Telescope) data bases*.[6,7] In the case of *Chandra*, all the data used were obtained with the Advanced CCD imaging spectrometer instrument on board the *Chandra X-ray Observatory*.[8] It consists of ten CCDs designed for efficient X-ray detection and spectroscopy. Four of them are front illuminated and organized in a square array with each CCD tipped slightly to better approximate the curved focal surface of the *Chandra* Wolter type I mirror assembly. The rest of the six CCDs are set in a linear array, tipped to approximate the Rowland circle of the objective gratings. Each CCD subtends an $8.4 \times 8.4$ arcmin$^2$ on the sky and the individual pixels of the CCDs subtend 0.492 arcsec on the sky (Garmire et al. 2003).

From the total sample of 18 blazars, 12 are reported in the *Chandra* Data Archive (see Table 1). In particular, we looked for observations that fulfilled the requirements needed for the analysis of the variability in this band. Namely, we looked for observations with a considerable number of counts (or *events*) and exposure times longer than 20 ks. In the case of the data used, these exposure times spanned from 27 to 105 ks, and the number of event counts spanned between 32 000 and 690 000. The energy band covers a range of 0.1–10 keV. All the data were reduced with the CIAO software packages version 4.7 with CALDB 4.6.2.

Regarding *Swift*-XRT, this telescope is a sensitive, flexible, autonomous X-ray CCD imaging spectrometer designed to measure the position, spectrum, and brightness of gamma-ray bursts over a range covering more than seven orders of magnitude in flux. The *Swift*-XRT data were processed using the most recent versions of the standard *Swift* tools: *Swift* software version 3.9, FTOOLS version 6.12, and XSPEC version 12.7.1. Light curves are generated using XRTGRBLC version 1.6 (Stroh & Falcone 2013). In total, 16 out of the 18 blazars were observed with *Swift*, but only 10 of them had observations with exposure times longer than 20 ks. The exposures times for these 10 sources range from 40 up to 520 ks and the energy band covers a range of 0.1–50 keV.

### 2.3 $\gamma$-ray data

The LAT on board the *Fermi* satellite,[9] is the $\gamma$-ray telescope with the largest collecting area and field of view up to date (Atwood et al.

---
[5] http://cda.harvard.edu/chaser/
[6] https://www.swift.psu.edu/monitoring
[7] https://www.swift.ac.uk/LSXPS/
[8] https://chandra.harvard.edu/

[9] https://fermi.gsfc.nasa.gov/

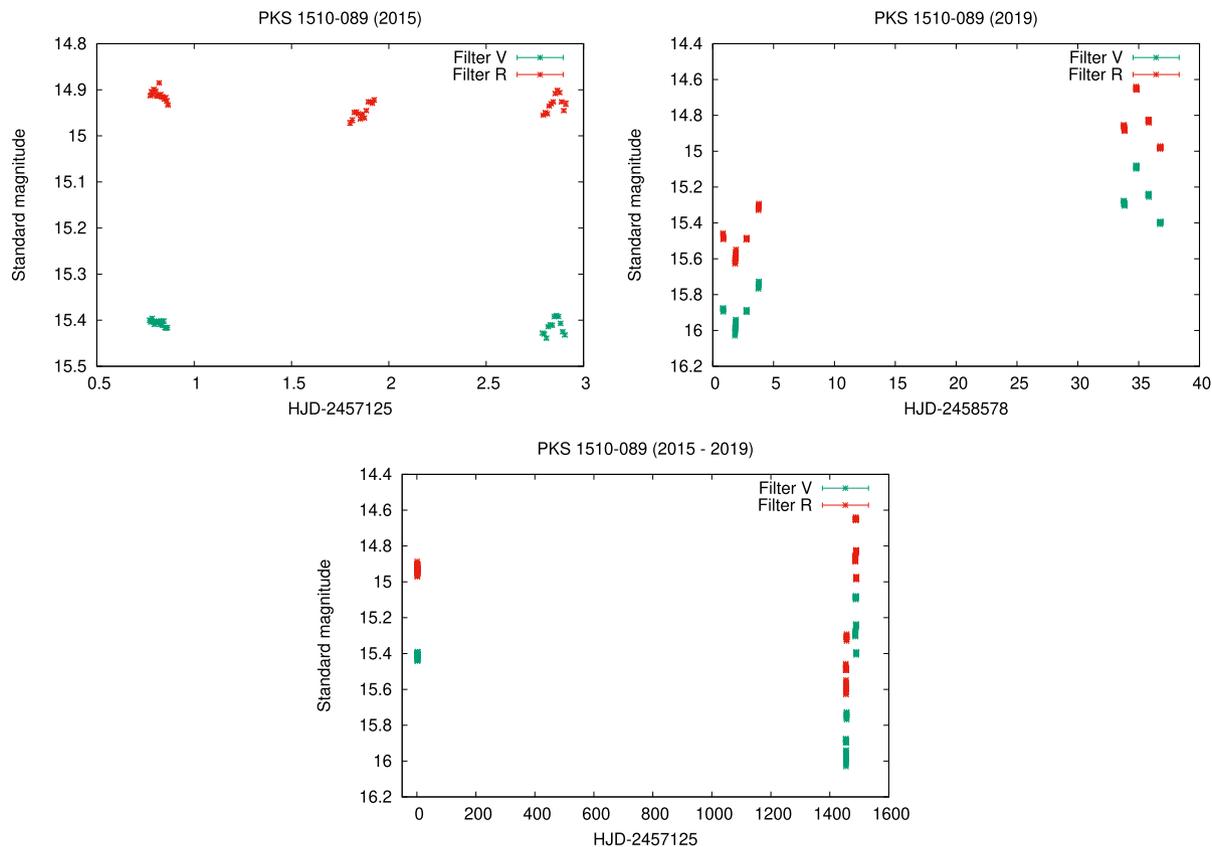

**Figure 2.** Standard magnitudes light curves in the *V* and *R* filters for PKS 1510−089, for 2015 (upper left panel), for 2019 (upper right panel), and the whole period 2015–2019 (lower panel).

2009), covering the energy band between 20 MeV and 300 GeV. This means it provides the most precise positional accuracy in its band, which ranges between 0.5 and 10 arcmin, depending on the source detection significance. This positional uncertainty is the reason behind the fact that ∼ 30 per cent of all sources in its latest catalog, the fourth *Fermi*-LAT Source Catalogue (Abdollahi et al. 2020) remain unidentified. In this work, we used the data taken from the third *Fermi*-LAT Source Catalogue (Acero et al. 2015), in order to generate the light curves.

From the total sample of 18 sources, only three blazars were not detected with *Fermi* ([HB89] 0414+009, PKS 2126−158, and PMN J2310−4374). All the data analysed here are available in the public data base archive.[10] The data-reduction process was made using the version v10r0p5 of the Fermi Science Tools.[11] The time range of the observations spanned from 2008 August to 2019 December, seeking the simultaneity of the optical data with those of the γ-ray band, as mentioned in Section 1.

## 3 ANALYSIS TOOLS

### 3.1 Variability

To generate the differential light curves (DLCs) for the analysis of the optical variability, we performed differential photometry, following Howell & Jacoby (1986). They used the source of interest (in this case, the blazar), and comparison and control stars (stars from the field). As a result, we obtained the DLCs, corresponding to the 'object-comparison' and 'control-comparison' curves. We applied the scale factor Γ (Howell, Mitchell & Warnock 1988, see their equation 13) which takes into account the differences in magnitude between the source and the comparison and control stars. Zibecchi et al. (2017, 2020) give a detailed analysis of the statistical results and how they are modified when Howell's factor is not considered. Optical variability results are shown in Table 3. The value of the dispersion σ for each DLC (i.e. $\sigma_1$ blazar-comparison and $\sigma_2$ control-comparison) were extracted for each DLC and used to perform the statistical analysis using the *C* criterion and the *F* test in order to study the source variability (see e.g. Zibecchi et al. 2017, 2020). We adopted a confidence level of 99.5 per cent for both statistical tests, being $C_{\rm crit} = 2.576$ the critical value for the *C* criterion and $F\text{-area}_{\rm crit} = 0.995$ for the *F* test.

In the case of the X-ray light curves, we used two methods: the GLVARY algorithm (only for the *Chandra* data) and the LCSTATS tool (for the *Swift* one) from NASA's HEASARC software. The GLVARY tool uses the Gregory–Loredo algorithm (Gregory & Loredo 1992), which splits the events into multiple time bins and looks for significant deviations between them. The tool assigns a variability index based on the total odds ratio, Odd; the corresponding probability of a variable signal, *P*; and the fractions of the light curve which are within 3σ and 5σ of the average count rate, $f3$ and $f5$. The possible values for the variability index are presented in Table 2. The other method, the LCSTATS tool, analyses how much the light curve deviates from a constant (or probability of constancy), performing statistical analysis for the time-series through the reduced $\chi^2$ and the Kolmogorov–Smirnov (KS) tests. Depending on the values of the constant source probabilities associated to both tests, the light curve is considered variable.

---

[10] https://fermi.gsfc.nasa.gov/cgi-bin/ssc/LAT/LATDataQuery.cgi
[11] https://fermi.gsfc.nasa.gov/ssc/data/analysis/scitools/

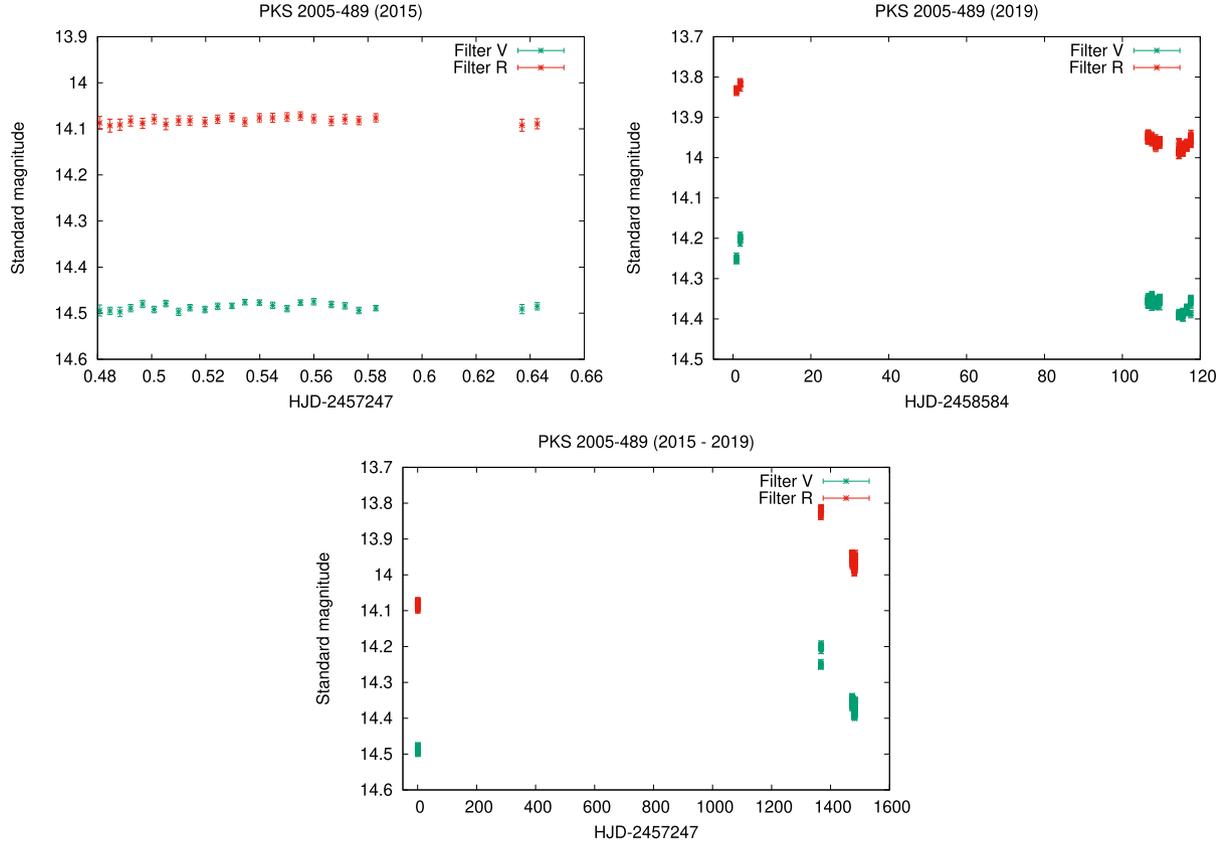

**Figure 3.** Standard magnitudes light curves in the *V* and *R* filters for PKS 2005−489, for 2015 (upper left panel), for 2019 (upper right panel), and the whole period 2015–2019 (lower panel).

On the other hand, since blazars dominate the high-energy sky (see e.g. Abdo et al. 2010a; Acero et al. 2015), several variability analyses have been performed on the *Fermi* data. In the first instance, we used the variability index published in the 3FGL catalogue. This index is calculated on light curves binned in intervals of one month, and constructed from the value of the likelihood of the null hypothesis that the source flux is constant over the full period of 4 yr. Associating the variability index to a distribution of $\chi^2$ with 47 degrees of freedom, if the index is greater than 72.44, the variability is considered probable with a 99percnt of confidence. As a complement, in order to make a deep analysis of the variability, we chose to use the *Fermi* All-sky Variability Analysis (FAVA) tool (Abdollahi et al. 2017) available online,[12] which generates light curves on demand for any given position in the sky within the *Fermi* footprint. The FAV analysis is independent from any diffuse emission model, and thus is not necessarily equal to the variability analysis given by the *Fermi* catalogues. In particular, FAVA can detect variability in unexplored areas of the sky given its inexpensive approach, does not depend on the spectral shape of any given source, and yields weekly binned light curves instead of the monthly binned results reported in the catalogues. Being a photometric approach, the only caveat with its use is that source location should be handled with care, since the LAT point spread function is relatively large (Ackermann et al. 2013, i.e. up to 10°). We obtained light curves for each sky position of our *Fermi* sources, which were determined with radio observations, and taking into account their corresponding 95 per cent confidence positional uncertainty ellipse, to avoid any possible source confusion.

### 3.2 Colour behaviour

#### 3.2.1 Magnitude–magnitude diagrams and the discrete correlation function

We investigated if any correlation between variations in *V* and *R* bands exists as well as the time lag between emissions. For this purpose, we used the discrete correlation function (DCF) applied to the magnitude–magnitude diagrams. This method was developed by Edelson & Krolik (1988), specially designed to analyse unevenly sampled data sets. All the equations used for computing the values of the DCF, as well as the errors, are well described in the paper mentioned before.

The usual convention is that a positive peak implies a correlation meanwhile a negative one corresponds to an anticorrelation. The height of the peak is a measure of how strong is the correlation. In our case, we expect at least a slight correlation, since the *V* and *R* emissions in the optical band are not so far apart within the electromagnetic spectrum. Depending on the obtained DCF value, we can infer which emission precedes the other. For example, if we had a positive correlation at a positive time delay, it imply that the variations in the *R* band precede the variations in the *V* band.

---

[12] https://fermi.gsfc.nasa.gov/ssc/data/access/lat/FAVA/

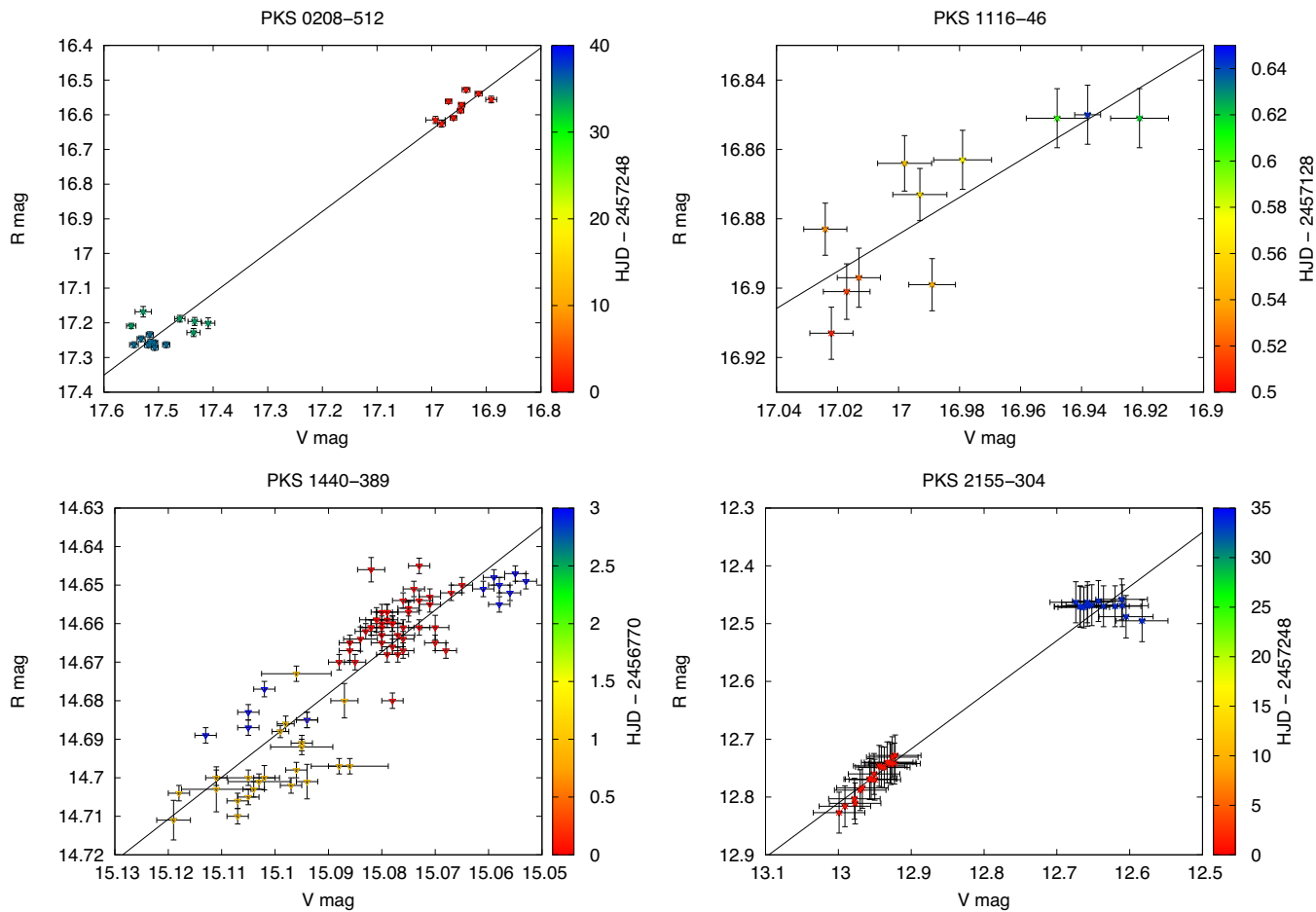

**Figure 4.** The magnitude–magnitude diagrams for PKS 1116−46, PKS 1440−389, and PKS 2155−304. The colours are ordered in time, according to the palette panel on the right.

### 3.2.2 Colour–magnitude diagrams

Since variations in the optical flux are usually related to spectral changes, which can be characterized by colour changes, the study of colour–magnitude diagrams may be helpful to find an explanation for the origin of variability in blazars. The analysis of the spectral changes can provide information about the physical processes that are responsible for these variations. In this work, we study the correlation between changes in the $(V - R)$ colour index against variations in the blazar's flux in the $V$ band. For each diagram, we calculated the corresponding Pearson's correlation coefficient, $r$, to quantify any possible trend. This coefficient is based on a lineal correlation between variables. A positive slope between colour index and magnitude implies a positive correlation, meaning that the blazar's trend is BWB. On the contrary, a negative slope evidences an opposite correlation, i.e. the source follows an RWB behaviour. In particular, we will define a correlation as *weak* when $|r| \leq 0.4$, *moderate* when $0.4 < |r| \leq 0.7$ and *strong* when $|r| > 0.7$.

As a complement, we took the average values of the $V - R$ colour index and we estimated the average spectral index (Wierzcholska et al. 2015):

$$\langle \alpha_{VR} \rangle = \frac{0.4 \langle V - R \rangle}{\log(\lambda_R / \lambda_V)}, \quad (1)$$

where $\lambda_V$ and $\lambda_R$ are the effective wavelengths in the $V$ and $R$ bands, respectively (Bessell, Castelli & Plez 1998).

## 4 ANALYSIS OF THE RESULTS AND DISCUSSION

### 4.1 Optical analysis

In Table 3, we present the resulting values of the $C$ and $F$ parameters in the optical band, the variability state, the number of points in the light curves, the $\Gamma$ factor value, the dispersion $\sigma_2$, the colour behaviour, the mean values of $V$ and $R$ magnitudes and their amplitudes and, finally, the spectral index $\alpha_{VR}$, for the total sample of 18 blazars. Taking into account both values of the statistical tools and applying the DLCs criteria proposed in Zibecchi et al. (2017, 2020), 27.7 per cent (5 out of 18) of the sample showed microvariability in the $V$ filter, meanwhile this percentage decreased to 22.2 per cent (4 of 18) for the $R$ filter.

On the other hand, a sub-sample of 11 blazars was observed in more than one night, within the corresponding observation runs. In our case, we considered the internight variability to that associated to time-scales of days, regardless of whether the days are consecutive or not. A 27.3 per cent of this sub-sample presented internight variability, in both $V$ and $R$ filters. On the other hand, we also observed 5 of the 18 blazars in a time-scale of months, finding variability in 4 of them. And for 3 of the blazars in the sample, we observed a significant variation in time-scales of years. The details are shown in Section 4.1.1.

From the analysis of the magnitude–magnitude diagrams and the DCF, we found that for all the sources where we detected microvariability as well as variations in time-scales of days and months, the value of the DCF resulted positive, which indicates a correlation between the emission in the *V* and *R* bands. I.e. the variations detected in both bands are produced by photons emitted in the same region and through the same physical process.

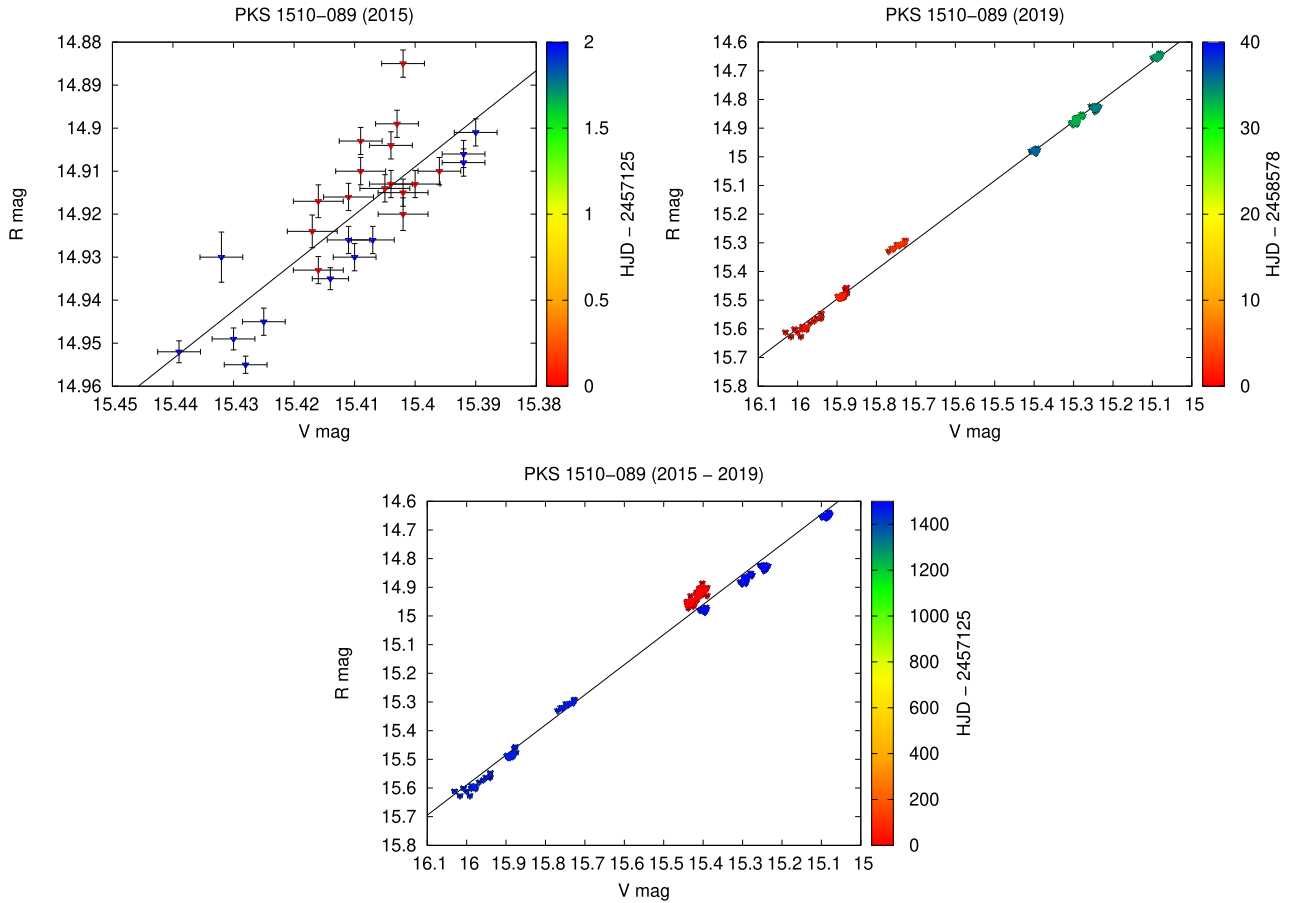

**Figure 5.** The magnitude–magnitude diagrams for PKS 1510−089 for 2015 (upper left panel), for 2019 (upper right panel), and the whole period 2015–2019 (lower panel). The colours are ordered in time, according to the palette panel on the right.

With respect to the colour–magnitude results, and taking into account the six blazars that presented variability in each of the time-scales considered, we found that all the objects reported a BWB trend on intraday time-scales (i.e. microvariability). All these objects presented a weak correlation with characteristics of the BWB type. Additionally, in one of the two nights where PKS 2155−304 presented variability, an RWB behaviour was detected. When considering interday time-scales (days, weeks, and months), in half of the sample (three out of six) a BWB tendency was found, split as one FSRQ (PKS 1116−46), and the other two BL Lacs (PKS 2005−489 and PKS 2155−304). In the case of the three blazars with an RWB trend, one corresponds to a BZU (PKS 0208−512), another one is a BL Lac (PKS 1440−389) and the last one (PKS 1510−089) is an FSRQ. Finally, when a year time-scale is considered, only two objects remained, PKS 1510−089 (FSRQ) with an RWB trend and PKS 2005−489 (BL Lac) with a BWB behaviour. All these results are in good agreement with the results obtained from Ikejiri et al. (2011) and Wang, Xiong & Bai (2019), where they claimed that most BL Lacs have strong BWB trends, and from (Gu et al. 2006b; Rani et al. 2010), where an RWB tendency is associated to FSRQs. In addition, we analyse the existence of hysteresis loops in the colour–magnitude diagrams. These loops illustrate the cyclical relationship between brightness and spectral changes during significant variations such as flares or outburst (e.g. Agarwal et al. 2021). For this purpose, we have included a palette in the colour–magnitude diagrams, which represents the observation date of each data point (see Figs 7–10). We have studied the cases where significant variations are reported, in particular for the blazars PKS 1510–089 and PKS 2005–512. In general, we did not find any loop-like behaviour in the colour–magnitude diagrams.

*4.1.1 Notes on individual objects*

In this section, we present a description for the behaviour of the sources that reported variability on some time-scale. In brief, we found that a subsample of six sources show variability, with PKS 0208−512, PKS 1116−46, PKS 1440−389, PKS 1510−089, PKS 2005−489, and PKS 2155−304 showing microvariability in at least, one of both filters (*V* and/or *R*). From this subsample, PKS 1116−46, PKS 1440−389, PKS 1510−089, and PKS 2005−489 presented significant variations in time-scales of days. Only in PKS 0208−512, PKS 1510−089, and PKS 2155−304, we detected variability in time-scales of months, with years-scale variations in the first two. It is worth noting that the scales of variability longer than hours depend on the time-scales on which data are available. Light curves are shown in Fig. 1 for the sources mentioned above showing variability at different time-scales.

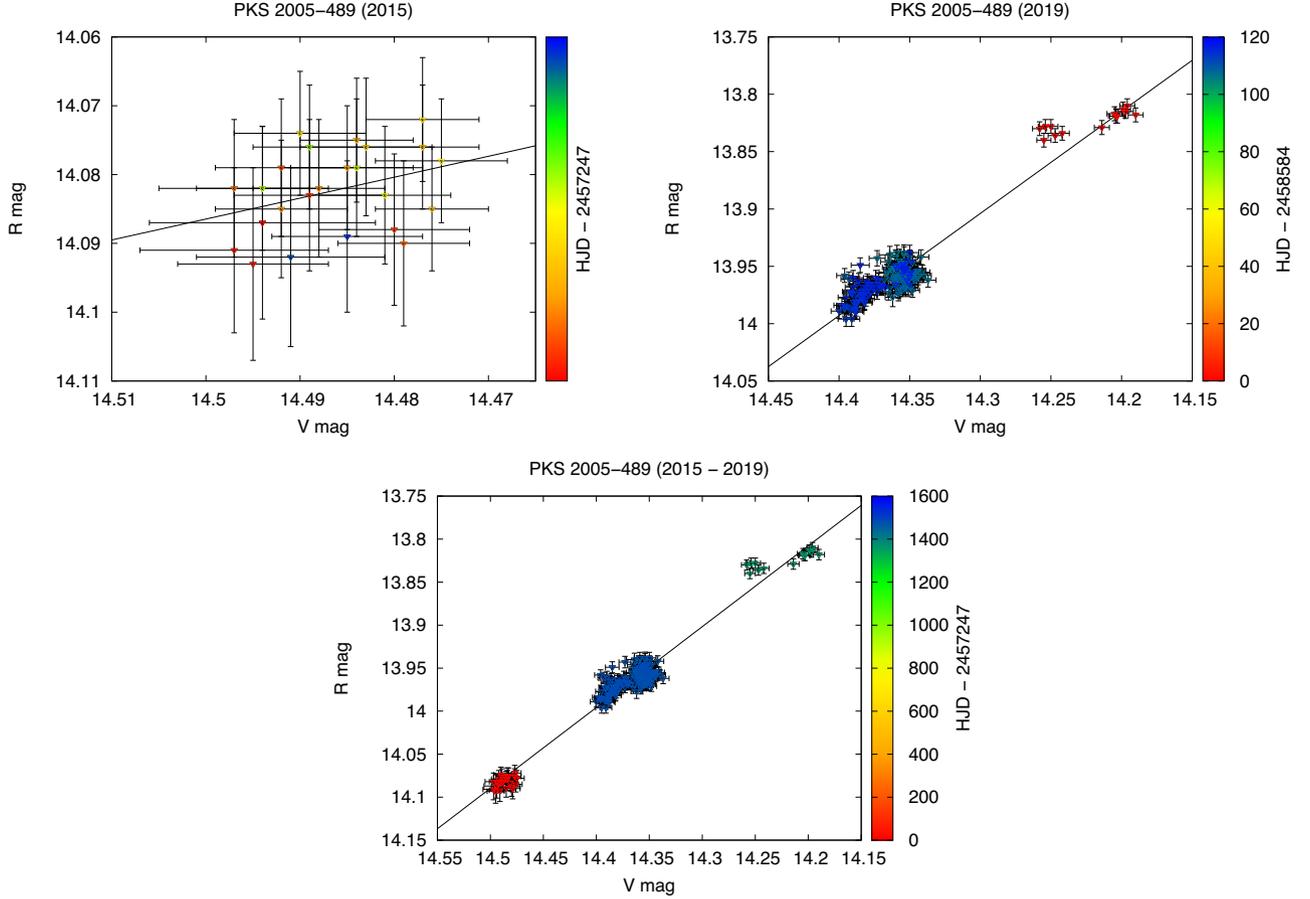

**Figure 6.** The magnitude–magnitude diagrams for PKS 2005−489 for 2015 (upper left panel), for 2019 (upper right panel), and the whole period 2015–2019 (lower panel). The colours are ordered in time, according to the palette panel on the right.

In the following, we provide an individual description for each of these objects, we analyse the behaviour of the detected variability and we study the possible correlations between magnitudes and $(V - R)$ colour. The DCF values given in each case correspond to the highest peak of the function, associated with a time lag of $\tau = 0$ s, which means that the variability, if exists, is affecting both the $V$ and $R$ bands at the same time. We adopted a 99.9 per cent confidence level for the Pearson coefficient.

PKS 0208−512: this blazar is catalogued as BZU, and it has displayed both long-term flux variations with high amplitude as well as significant microvariability. Its 2008–2012 $R$-band light curve shown in Chatterjee et al. (2013a) displayed a $\gtrsim 3$ mag amplitude. On the other hand, this source also reported a significant microvariation of $\Delta m_V = 0.131$ mag in 7.75 h, with a mean value of $V = 15.63 \pm 0.02$ mag (Romero et al. 2002).

We observed this blazar during three nights, 2015 August 13 and September 15 and 17 (see Fig. 1). Results can be seen in Table 3. For the 2015 data, we found microvariability in the $R$ band and significant variations on time-scales of months, in both $V$ and $R$ bands. The mean magnitude during the 2015 September 17 night ($\langle V \rangle = 17.26$ mag) indicates that the blazar was at a low-flux stage during our observations, with respect to the values mentioned above. With respect to the spectral index $\alpha_{VR}$ (see Table 3), the spectral distribution became harder during the period August–September (Fig. 11). We also obtained a strong correlation between $V$ and $R$ bands, with a value of the Pearson's correlation coefficient $r = 0.992$, and DCF $= 1.006 \pm 0.103$; this would indicate that $V$ and $R$ emissions are co-spatial (Fig. 4). Finally, analysing the colour–magnitude diagram (Fig. 7), although we found a weak BWB tendency for the night that the source resulted variable, a strong negative correlation is evident in time-scales of months, with $r = -0.759$, indicating an RWB tendency. The latter trend is similar to what Chatterjee et al. (2013a) reported.

PKS 1116−46: this source is an FSRQ, observed during two nights in 2015 April. We found intranight variations in both $V$ and $R$ filters, as well as at internight time-scales in the $R$ band (Table 3). This is the first time that microvariability is reported for this source. In Fig. 1, we show the magnitude variations for both filters. We found similar values for the standard magnitudes to those reported in the literature (Tritton 1971; Adam 1985; Bozyan, Hemenway & Argue 1990; Ojha et al. 2009). This source has maintained an almost constant $m_v$ value over the last 45 yr approximately, becoming slightly brighter lately. Magnitudes in both filters presented a strong correlation (Fig. 4), supported by the value of the DCF $= 0.889 \pm 0.11$ and the corresponding Pearson's correlation coefficient $r = 0.829$. About the colour behaviour, the colour–magnitude diagram (Fig. 7) showed a moderate BWB tendency for the blazar, with a Pearson's coefficient $r = 0.541$.

PKS 1440−389: we observed this BL Lac (HSP) during three nights in 2014 April. Both internight and intranight variabilities were detected, but only in $V$ filter (Table 3 and Fig. 1). We found a strong correlation between the flux in both bands (Fig. 4), where the values of the DFC and the Pearson's coefficient confirmed this (DCF $= 0.053 \pm 0.030$ and $r = 0.877$). From the colour–magnitude diagram (Fig. 7), we obtained that the blazar could present an RWB tendency, with a weak and negative correlation for the data (Pearson's coefficient $r = -0.176$).

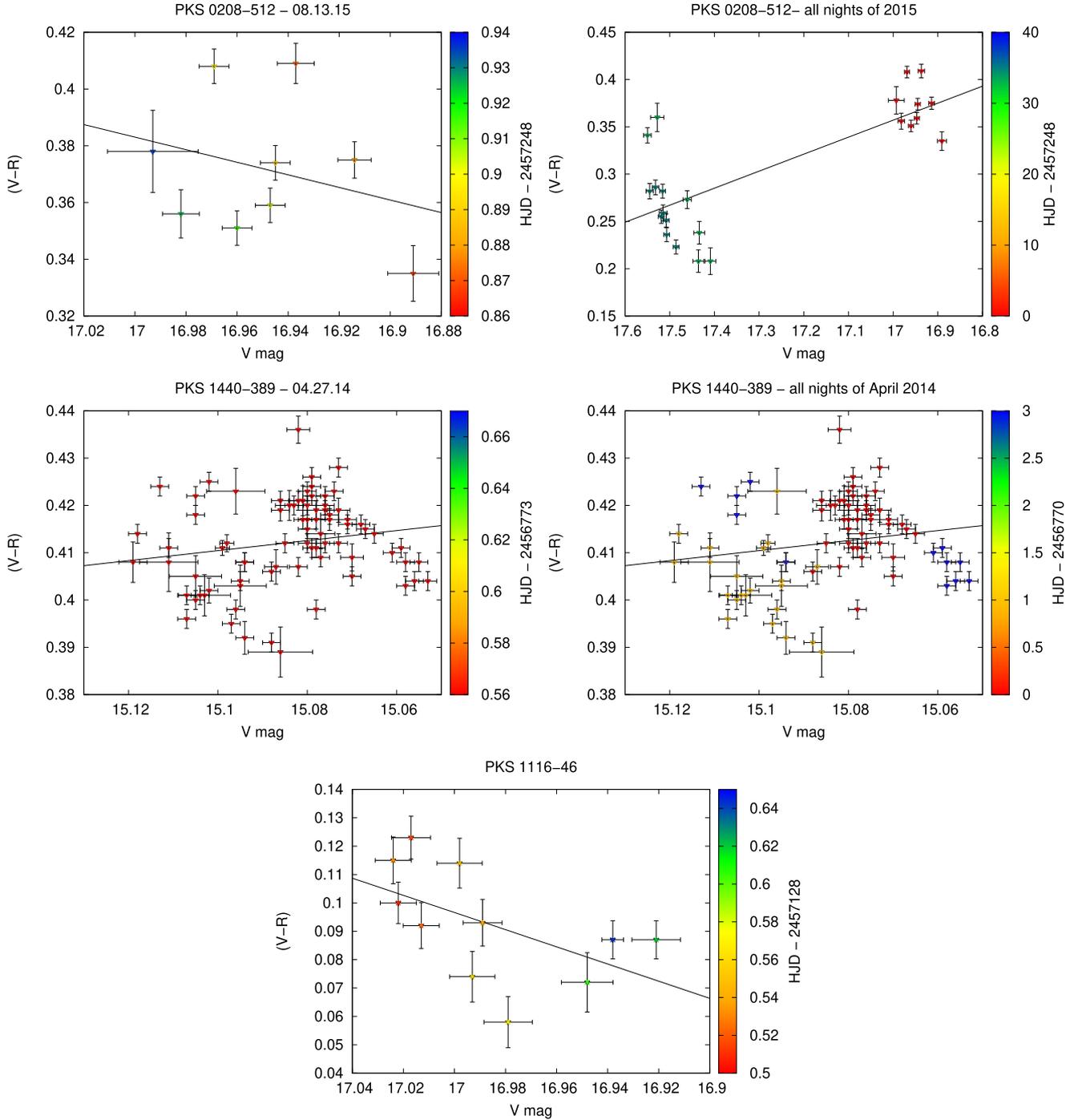

**Figure 7.** The colour–magnitude diagrams for PKS 0208−512, PKS 1116−46, and PKS 1440−389. The colours are ordered in time, according to the palette panel on the right.

PKS 1510−089: This is a well-studied FSRQ, observed during 11 nights, between 2015 April 12–14, 2019 April 4–7, and 2019 May 7–10. From the 2015 data, intranight variability was detected during the third night, in both filters. With respect to the internight variability, we found some variations, specially in the *R* filter (Table 3). In 2019, we found microvariability in three nights in the *R* filter, while for the observations of April and May, internight variations were detected in both filters (Fig. 2). Gupta et al. (2016) reported the behaviour of this blazar in time-scales of months (from 2014 April–August), obtaining amplitudes $\Delta m_V = 0.85$ mag and $\Delta m_R = 0.75$ mag. Comparing its magnitudes between 2014 and 2015, PKS 1510−089 was increasing in brightness (0.85 mag for *V* and 0.87 mag for *R*). Sandrinelli et al. (2014) also studied the variability in short and long timescales between 2006 January 2006 and 2012. They reported this source as strongly variable, with amplitudes of $\Delta m_V = 3.25$ mag and $\Delta m_R = 3.01$ mag in ∼6.5 yr. As can be seen from the data provided in this paper, we also find this object as a variable blazar. From the results obtained in this work together with those published by the Yale/SMARTS[13] blazar monitoring group, we studied the behaviour of $\alpha_{VR}$ in long time-scales. Based on the light curve variations, we analysed the regions where a magnitude increment took place, and calculated the corresponding values of $(V − R)$ and $\alpha_{VR}$. In Fig. 13, we show the variations in the $V$ and $R$ fluxes and the corresponding behaviour of $\alpha_{VR}$. When the blazar became weaker, the value of the $\alpha_{VR}$ index decreased, implying that the spectral energy distribution became harder. Instead, when the source was brighter, the spectral index increased, i.e. the spectral distribution got softer. Thus, the changes in the brightness of the blazar were reflected in the spectral behaviour, as expected. Due to this behaviour, we also expected a correlation between the *V* and *R* bands, as can be seen in the magnitude-magnitude diagram in Fig. 5. We found that both the DCF value (DCF = $0.728 \pm 0.080$) and the Pearson's coefficient ($r = 0.816$) corroborated the strong positive correlation (see Section 4.2 for a detail analysis). For the other set of data (2019), similar results were obtained, with a DCF value of DCF = $0.995 \pm 0.085$ and a value of $r = 0.999$, giving a strong positive correlation between *V* and *R* emissions. With respect to the colour–magnitude diagram (Fig. 8), we obtained a weak negative correlation (with a Pearson's coefficient of $r = -0.146$), implying a possible RWB tendency in 2015. In 2019, a moderate RWB tendency was obtained, with a Pearson's coefficient $r = -0.616$. Finally, taking into account the time-scale of years, from 2015 to 2019, we found a weak negative (moderate) correlation ($r = -0.400$), indicating an RWB trend.

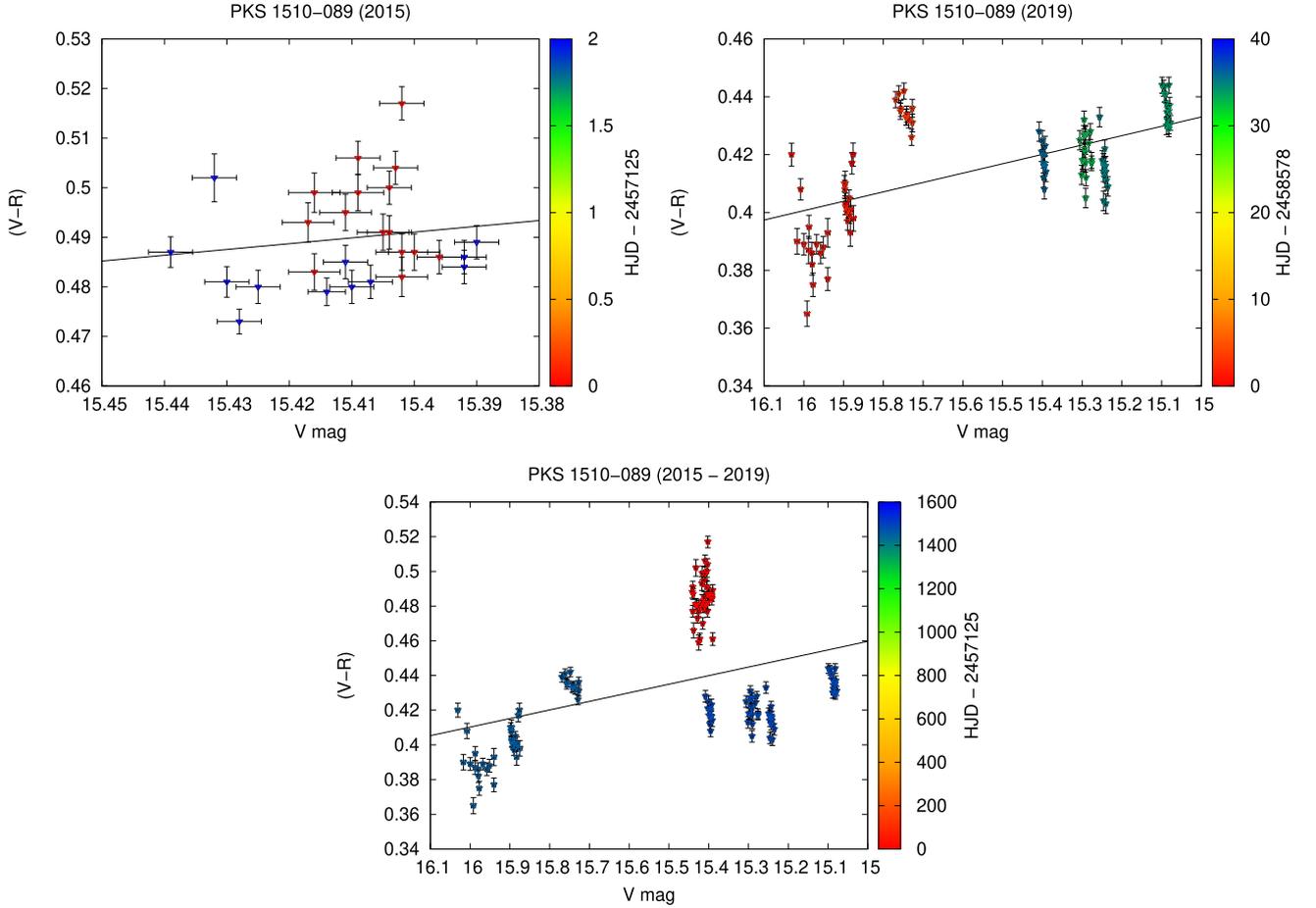

**Figure 8.** The colour–magnitude diagrams for PKS 1510−089 for 2015 (upper left panel), for 2019 (upper right panel), and the whole period 2015–2019 (lower panel). The colours are ordered in time, according to the palette panel on the right.

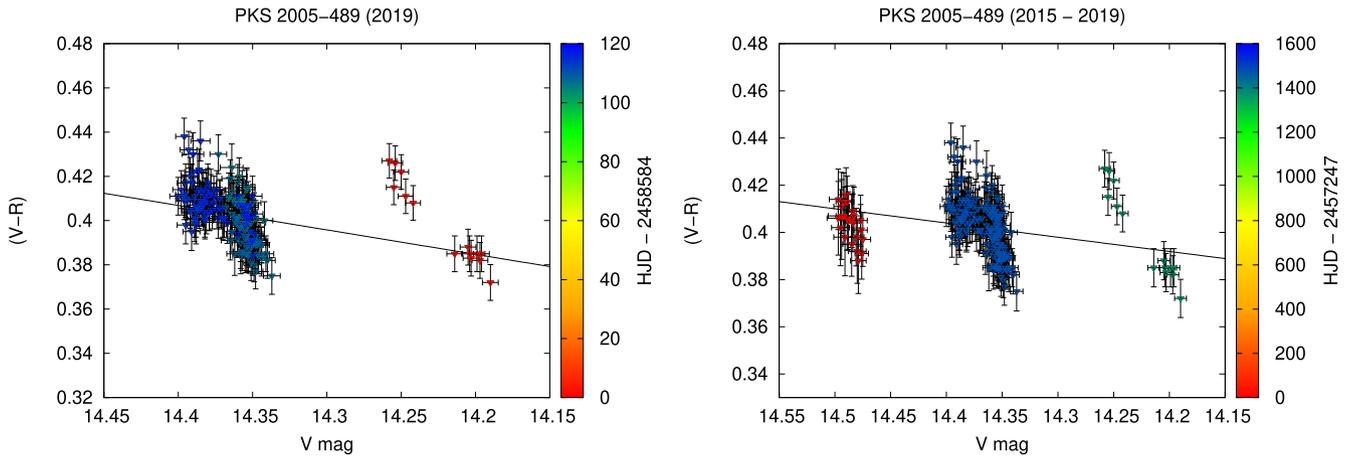

**Figure 9.** The colour–magnitude diagrams for PKS 2005−489 for 2015 (upper left panel), for 2019 (upper right panel), and the whole period 2015–2019 (lower panel). The colours are ordered in time, according to the palette panel on the right.

[13] www.astro.yale.edu/smarts/glast/home.php

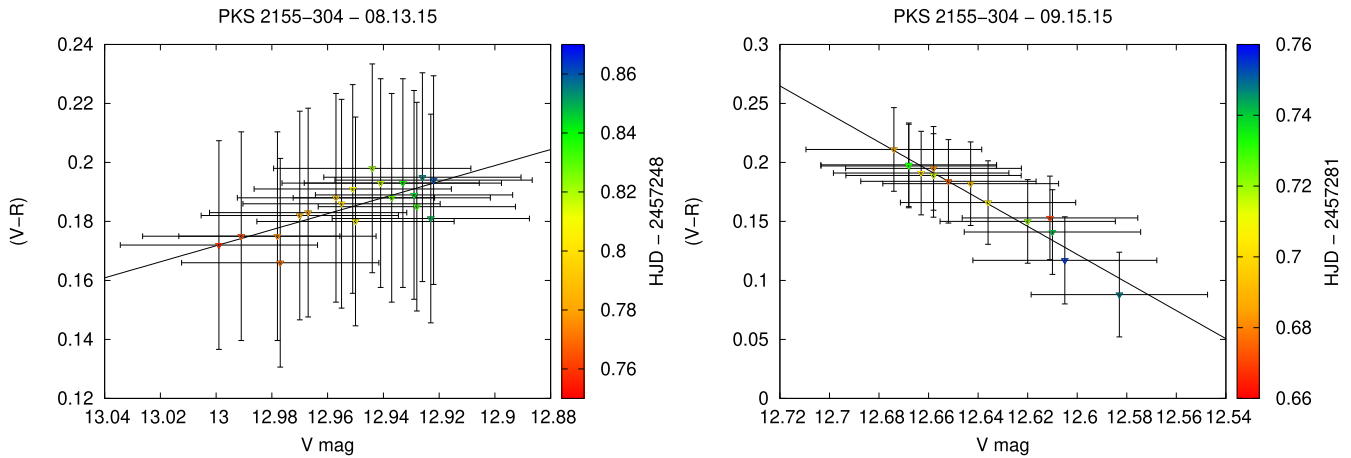

**Figure 10.** The colour–magnitude diagrams for PKS 2155−304, for the two data sets of 2015 August and September. The colours are ordered in time, according to the palette panel on the right.

PKS 2005–489: This BL Lac (HSP) was observed during 11 nights spread from 2015 August 2015 to 2019 September . No intranight variability was found neither in 2015 nor in 2019. On the other hand, variability in time-scales of months and years was detected for both bands, *V* and *R* (Fig. 3 and Table 3). Heidt & Wagner (1996) observed this blazar in June 1990 and found a mean value of 13.5 mag for the *R* filter, meanwhile Mahony et al. (2011) reported a value of 14.55 mag in the same filter for their observations carried out in 2011 October. Sandrinelli et al. (2014) obtained similar results to ours. The magnitude–magnitude diagram (Fig. 6) showed a weak correlation between the *V* and *R* bands, supported by the values of DCF = $0.285 \pm 0.084$ and the Pearson's coefficient $r = 0.344$ (2015), and DCF = $0.296 \pm 0.075$ and $r = 0.957$ (2019). We respect to the colour–magnitude correlation, for the 2015 data, the amplitude of both magnitude and colour variations was within the errors, so the colour–magnitude trend is spurious. For 2019, we obtained a weak positive correlation ($r = 0.322$), implying a possible BWB trend, and considering the whole period (2015–2019), we found a weak BWB tendency (with a $r = 0.287$). In these cases, both tendencies found are affected by correlated errors (Fig. 9).

PKS 2155–304: we observed this very well-known BL Lac (HSP) during two nights in 2015. We detected intranight variations in both nights, in both the *V* and *R* bands (Table 3). We also detected variations with a significant amplitude between August and September in both filters, finding that the source became brighter by almost half a magnitude in barely more than a month. We show the light curves in Fig. 1. Several authors (Griffiths et al. 1979; Miller & McAlister 1983; Hamuy & Maza 1987; Carini & Miller 1992; Smith et al. 1992; Jannuzi, Smith & Elston 1993; Heidt, Wagner & Wilhelm-Erkens 1997; Romero, Cellone & Combi 1999; Sandrinelli et al. 2014) studied the behaviour of the light curves of this blazar. Flux changes were similar in both bands, with the brightness of the blazar increasing during the period 1979–2000, and then decreasing between 2005 and 2015. In the 1995–2005 period, the blazar presented the highest level of variability activity. The analysis of the magnitude–magnitude diagram (Fig. 4) shows a strong correlation between magnitudes in both filters; this is supported by the values of DCF = $0.915 \pm 0.055$ and the Pearson's coefficient $r = 0.989$. From the published data of Hamuy & Maza (1987), Smith et al. (1992), and Sandrinelli et al. (2014) together with our values, we analysed the behaviour of $\alpha_{VR}$ during 30 yr, finding that the spectral index $\alpha_{VR}$ increased during the same time span as the flux in the *V* and *R* bands, while it decreased in line with the blazar's brightness decay. This behaviour implies that as the blazar became weaker, the spectral energy distribution became harder. Conversely, when the brightness of the blazar increased, its spectral distribution became softer. Regarding the colour–magnitude diagram shown in Fig. 10, we observed different behaviours in each data set: From the August data, we obtained a moderately negative correlation, with a Pearson's coefficient $r = -0.741$, implying an RWB tendency; and, for September data, we obtained a strong positive correlation ($r = 0.972$), associated with a BWB trend. The global trend seen for both data sets was a barely moderate positive correlation ($r = 0.406$), implying a slight BWB tendency.

The spectral variations for the blazars described in detail before are presented in the SEDs of Fig. 11.

### 4.2 Multiwavelength analysis

Our aim is to perform a multifrequency analysis focused on the possible correlation between optical, X- and/or $\gamma$-ray variabilities. In this sense, from the total sample of 18 blazars, 12 were observed with *Chandra*, and we only found variability for 5 of them in time-scales of hours. With respect to *Swift*, 16 out of 18 blazars had observations, and 9 showed significant variations in its light curve. On the other hand, a total of 15 blazars were included in the 3FGL catalogue and 8 of them reported $\gamma$-ray variability. Considering only the sources that showed variability in all optical, X-ray, and $\gamma$-ray bands, we have four sources: PKS 0208−512, PKS 1510−089, PKS 2005−489, and PKS 2155−304 (see Table 4). In particular, for PKS 1510−089 a series of important flares were detected at these frequencies. Based on the temporal coverage as well as the amount of data available in the three bands, we only performed a detailed multiwavelength analysis on PKS 1510−089 and PKS 2005−489. In Table 5, we summarize all the information for these two sources. For the other sources (PKS 0208−512 and PKS 2155−304), we are expecting new data in the following months, which will be published in a forthcoming paper.

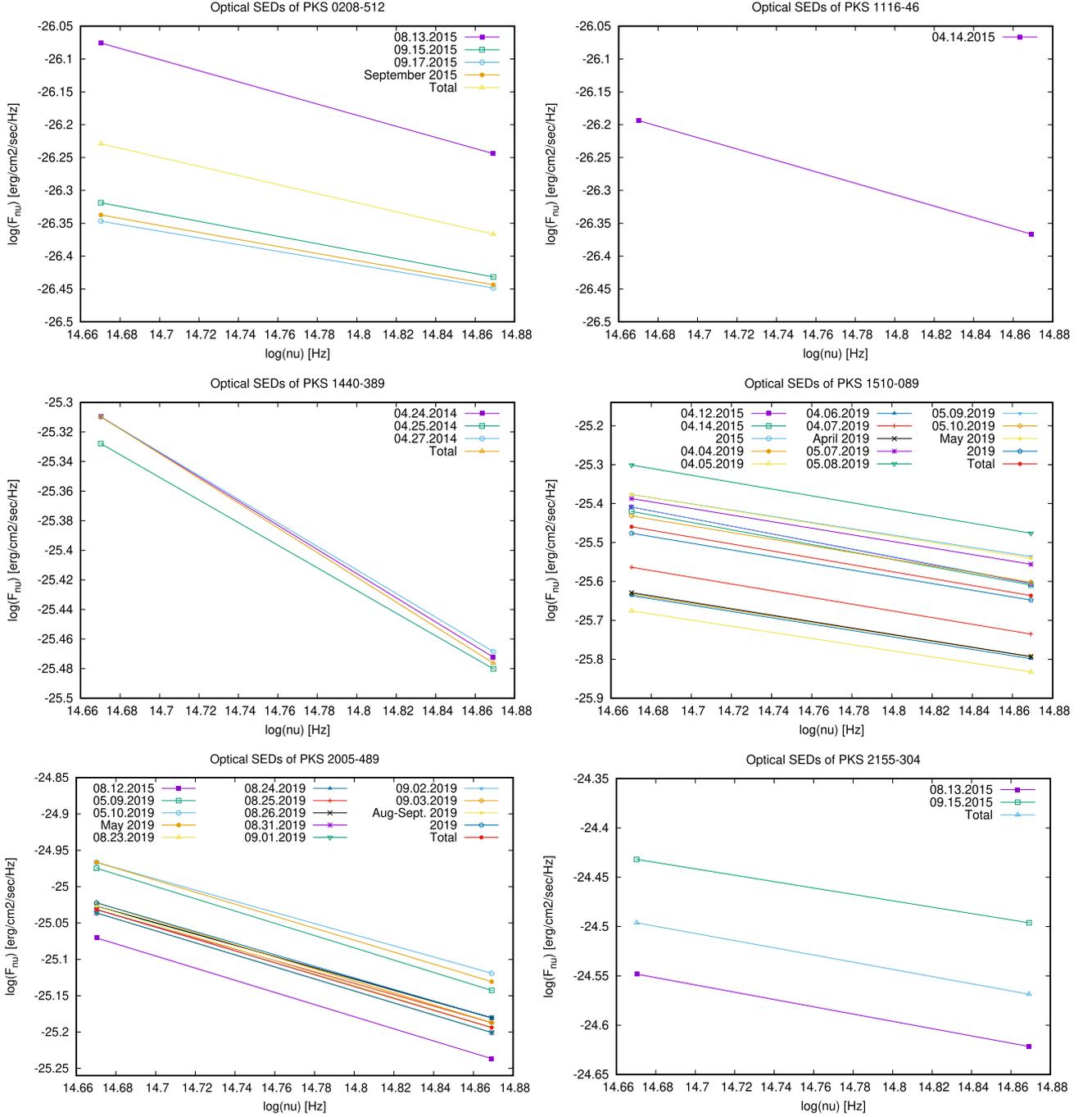

**Figure 11.** The SEDs corresponding to the blazars PKS 0208−512, PKS 1116−46, PKS 1440−389, PKS 1510−089, PKS 2005−489, and PKS 2155−304.

In the case of PKS 1510−089 (see Fig. 13), for the *Chandra* data and taking into account the whole observation run (12.6hr), we did not detect any variations in the X-ray light curve, obtaining a value of the GLVARY tool equal to 0 at this time-scale, meaning it is definitively non-variable (see the description of the GLVARY tool in Section 3.1). This result is corroborated with the $\chi^2$ test ($\chi^2 = 1.178$). The *Swift*-XRT data allow us to study the X-ray variability at time-scale of years. The results of the LCSTATS tool showed a significant variability, with a value of the reduced $\chi^2_{\rm red} = 918.4$ (associated with a $\chi^2 = 469340$ and the degrees of freedom d.o.f = 511) and probabilities of constancy of $\text{Prob}(\chi^2) = 0$ and $\text{Prob}(KS) = 0$. On the other hand, according to the 3FGL catalogue, its $\gamma$-ray variability index is 11 014.0, well above the associated critical value of 72.44, indicating a significant variability in the $\gamma$-ray band. There are five activity phases in the $\gamma$-ray light curve: the first activity phase took place between 2008 September and 2009 June, the second one between 2011 April and 2012 July, the third one between 2013 June and 2014 February, the fourth one between 2014 August and 2015 November, and the last one between 2016 March and December. We define as a high-activity phase the observation run in which the flux reaches three times its median absolute deviation for all the data. All phases show brightness amplitudes well above four times the average flux level of this source. We also find there is a slight increment in brightness during the last time bins in the 3FGL light curve, hinting that a sixth phase could have been taking place at that moment. In order to corroborate this, we used the FAVA tool (Abdollahi et al. 2017) available online (see Fig. 13). The FAVA light curve shows activity immediately after the last period covered by 3FGL, albeit a very mild one: The detected flux during this period is only twice the average flux for this source.

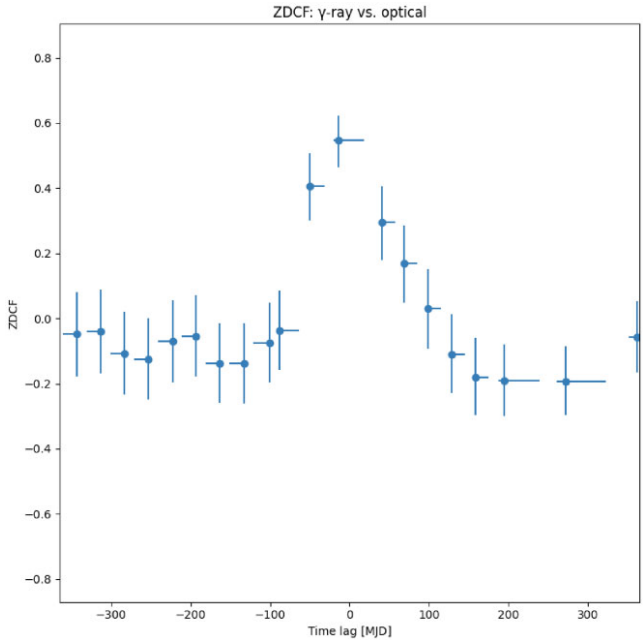

**Figure 12.** $Z$-transformed DCF between the $\gamma$-ray and optical ($V$ and $R$) data for the FSRQ PKS 1510−089. The correlation is maximum at a lag compatible with less than 20 d, with $\gamma$-rays preceding the optical emission. The errors are estimated by 10 000 Monte Carlo simulations.

The other variable blazar in our multiwavelength analysis is PKS 2005−489 (see Fig. 14). We found a significant variation in its X-rays light curve for the *Chandra* data, with a value of the GLVARY tool of 7 (definitively variable). The time-scale associated with these variations is approximately 10 d. With respect to the *Swift*-XRT light curve (Fig. 14), we obtained a value of the reduced $\chi^2_{\text{red}} = 1700$ (associated with a $\chi^2 = 869450$ and the degrees of freedom d.o.f = 511) and probabilities of constancy of $\text{Prob}(\chi^2) = 0$ and $\text{Prob}(KS) = 0$, from the LCSTATS tool. At the same time, its variability index given by the 3FGL catalogue for the $\gamma$-ray band is 131.057, greater than the associated critical value of 72.44, which means that the source is variable. However, the $\gamma$-rays light curve shows a much lower amplitude than that of PKS 1510−089: The maximum increment detected is 1.5 times its time-averaged flux. The flux curve for the *Fermi* data shows very mild variability in $\gamma$-ray flux contemporary to the time in which we detect variations in its X-ray flux.

## 5 DISCUSSION: COLOUR–BRIGHTNESS TENDENCY

We are interested in discussing the colour–brightness relation in blazars, and whether this relation can be seen within our data for the only two variable blazars in our multiwavelength sample, i.e. PKS 1510−089 and PKS 2005−489. In particular, an optical colour–intensity relation in blazars has been suggested by many authors, sometimes with contradictory claims.

A colour–intensity study was performed for a statistically significant sample of blazars by Ikejiri et al. (2011), on a sample of 42 sources. In that work, they claim that the BWB trend is a common characteristic in all blazars, with variations on time-scales of days/months. Moreover, they suggest that an RWB trend may be dominant when a BL Lac source undergoes a low activity phase, while BWB tendencies would be associated to high activity. The RWB behaviour could then be explained by the relatively stronger contribution of the (bluer) thermal accretion disc during low activity, with respect to the redder and more variable contribution of the jet. A BWB trend during high-activity phases, in turn, may be attributed to shock-in-jet processes within the jet, or to Doppler factor variation on a convex spectrum (Villata et al. 2004).

On the other hand, Safna et al. (2020) performed a similar analysis on a sample of 37 sources, in which they found the RWB to be the dominant trend instead, for FSRQs. They also found that flux variability and variability amplitude increase towards longer wavelengths for both classes of sources, which suggests that the jet emission dominates over the bluer, less bright thermal emission arising from the accretion disc.

However, it is worth noting that among the 42 sources reported by Ikejiri et al. (2011), 29 are BL Lac sources and 13 FSRQs. Given that four of the FSRQs were observed only five times or less, we consider that their results on FSRQs should be confirmed with observations on a larger sample. A similar case can be made for the BL Lacs studied by Safna et al. (2020), since four out of seven do not show the RWB trend seen in the majority of their sample. Thus, following these studies, it is possible that the BWB behaviour is typical of BL Lac sources, as suggested by Ikejiri et al. (2011), and the RWB is typical of FSRQ sources as suggested by Safna et al. (2020). The inverse remains to be confirmed or discarded. This is in line with past results: for example, BL Lacertae itself has shown a BWB behaviour (Villata et al. 2002, 2004) while the FSRQ object 3C 454.3 has shown an RWB trend (Villata et al. 2006). Gu et al. (2006b) found the same colour variation difference between both types of sources in a sample of 8 blazars. Similar trends were reported by Massaro et al. (1998), Vagnetti et al. (2003), Wu et al. (2011), and Bonning et al. (2012), to name a few. In particular, Zhang et al. (2015) also report an RWB trend in 35 out of 46 FSRQ sources, and a BWB trend for 11 out of 18 BL Lac objects.

We wish to discuss colour–brightness tendencies (detailed in Section 4.1.1), with the multiwavelength light curves (Section 4.2) and the optical variability (Section 4.1.1). To add data to the discussion, we included in Fig. 13 and in Fig. 14 our data alongside optical data available in the literature (taken from Bonning et al. 2012) for PKS 1510-089, and from the ATOM (Automated

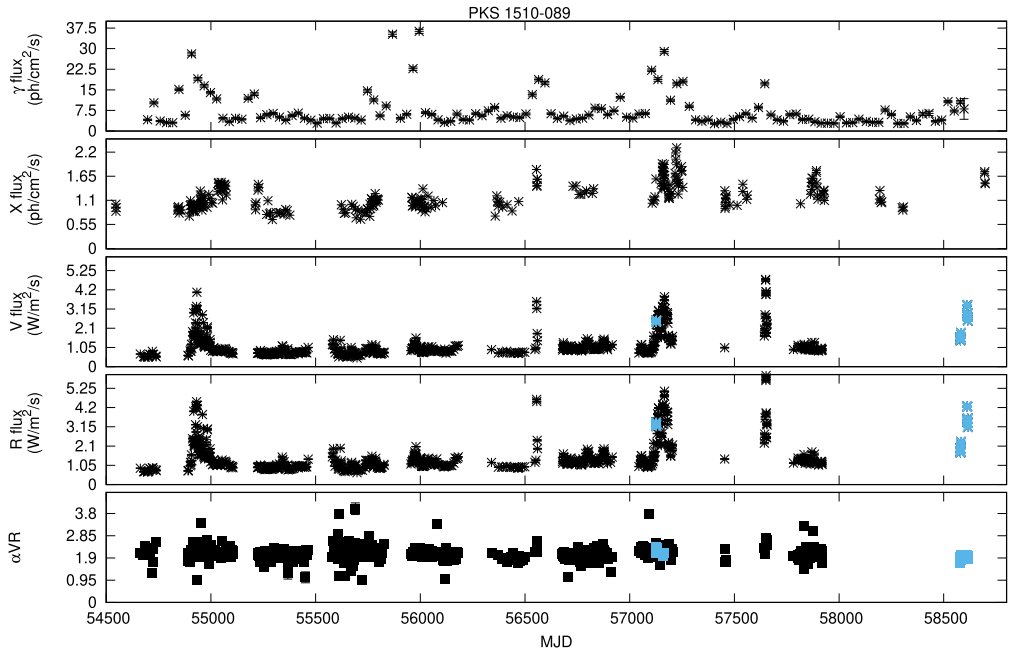

**Figure 13.** Multiwavelength light curves for the blazar PKS 1510−089 and its index $\alpha_{VR}$. The X-ray flux is in units of $10^{-11}$ ph cm$^{-2}$ s$^{-1}$. The $\gamma$-ray flux is in units of $10^{-7}$ ph cm$^{-2}$ s$^{-1}$.

**Table 4.** Variability states for the blazar sample in the optical, X−rays, and $\gamma$-rays.

| Object | Optical | X−rays | $\gamma$-rays |
| --- | --- | --- | --- |
| PKS 0208−512 | ✓ | ✓ | ✓ |
| [HB89] 0414+009 | ✗ | ✓ | ✗ |
| PKS 0521−36 | ✗ | ✓ | ✓ |
| 3FGL J0846.9−2336 | ✗ | ✗ | ✗ |
| PKS 1116−46 | ✓ | ✗ | ✗ |
| PKS 1127−14 | ✗ | ✓ | ✓ |
| PKS 1229−02 | ✗ | ✗ | ✗ |
| PMN J1256−1146 | ✗ | ✗ | ✗ |
| PKS 1424−41 | ✗ | ✓ | ✓ |
| PKS 1440−389 | ✓ | ✗ | ✗ |
| PKS 1510−089 | ✓ | ✓ | ✓ |
| 3FGL J1917.7−1921 | ✗ | ✗ | ✗ |
| 3FGL J1958.2−3011 | ✗ | ✗ | ✗ |
| PKS 2005−489 | ✓ | ✓ | ✓ |
| PKS 2126−158 | ✗ | ✗ | ✗ |
| PKS 2149−306 | ✗ | ✓ | ✓ |
| PKS 2155−304 | ✓ | ✓ | ✓ |
| PMN J2310−4374 | ✗ | ✗ | ✗ |

Telescope for Optical Monitoring of H.E.S.S.) Data Base[14] for PKS 2005−489.

In the case of the FSRQ source PKS 1510−089 (see Fig. 13), it is possible to detect at least five main activity phases in the Fermi light curve. Our optical data coincide with the fourth phase, and hint at a sixth phase that was beginning by the end of the last observation run. We are missing optical observations during the second high $\gamma$-ray activity phase. Our X-ray data, on the other hand, show no significant variability. Since PKS 1510-089 has its SED valley in the soft X-ray band (around 2 keV, see e.g. Barnacka et al. 2014), the

---

[14] https://www.lsw.uni-heidelberg.de/projects/hess/ATOM/

X-ray flux is rather low. Abdo et al. (2010d) also found complex microvariability (within 6–12 h) in the optical and $\gamma$-rays bands of PKS 1510−089. They demonstrated there was a strong correlation between the $\gamma$-ray and optical fluxes, with the latter leading the former in 13 d. Similar conclusions were claimed to be derived by Rajput, Stalin & Sahayanathan (2020), by cross-correlating the $\gamma$-ray and optical light curves for some specific, arbitrarily chosen periods. A detailed cross-correlation study was presented recently by Yuan et al. (2023), both for the whole period and for each single high-activity phase. In that study, they used the Z-transformed discrete correlation function (ZDCF), developed by Alexander (1997), which is adapted for sparse, unevenly sampled light curves, as is the case with most blazar observations. They found there is a strong correlation between the optical and $\gamma$-ray bands, with a time lag compatible with less than 10 d.

Moreover, the variations in the different fluxes from PKS 1510−089 seem to be related to changes in the colour behaviour. Sandrinelli et al. (2014) found an RWB trend during the low activity phases, while the $(R - H)$ and $(J - K)$ colours indicated a BWB trend during the high-activity phases. Similar results were found by Sasada et al. (2011), who also studied the same activity phase in 2009 (with a 10× flux increase in 10 d), and found an average RWB tendency when analysing the $(V - J)$ colour, but a BWB trend when the activity ensued. Analysing our optical data, PKS 1510−089 presented a moderately RWB ($r = -0.40$) overall averaged tendency, since it was undergoing a quiet phase during 2015. However, when taking into account only the time interval with both strong intranight and internight variability (2019), it displayed a moderate to strong RWB behaviour: $r = -0.62$ during April 2019, and $r = -0.82$ during 2019 May. Being an FSRQ, this result is in agreement with Sasada et al. (2011). However, the fact that both the $\gamma$-ray and optical bands, which are close to the inverse Compton and synchrotron peaks, respectively, are varying and could be correlated, means that the jet component is most likely responsible for this variability. This means that variations in the

**Table 5.** Summary of the results obtained for PKS 1510−089 and PKS 2005−489 in the multiwavelength analysis.

| Object | SED class | Optical | | CM trend | Pearson's coefficient $r$ | CI 95 per cent | X-rays 0.1–8 keV | $\gamma$-rays 20 MeV–300 GeV |
| --- | --- | --- | --- | --- | --- | --- | --- | --- |
| | | $V$ | $R$ | | | | | |
| PKS 1510−089 (2015) | FSRQ (LSP) | V | V | RWB | −0.15 (weak) | [−0.4, 0.0] | V | V |
| (2019) | | V | V | RWB | −0.62 (moderate) | [−0.7, −0.5] | V | NV |
| (2015-2019) | | V | V | RWB | −0.40 (weak) | [−0.7, −0.5] | V | V |
| PKS 2005−489 (2015) | BL Lac (HSP) | NV | NV | – | – | – | V | V |
| (2019) | | V | V | BWB | 0.33 (weak) | [0.2, 0.4] | NV | NV |
| (2015–2019) | | V | V | BWB | 0.29 (weak) | [0.2, 0.4] | V | V |

*Notes.* The columns are object name, SED class, variability state in the optical band, the Pearson's coefficient $r$, the colour–magnitude trend, the confidence interval (CI) at 95 per cent of confidence level, and the variability state in the X- and $\gamma$-rays bands.

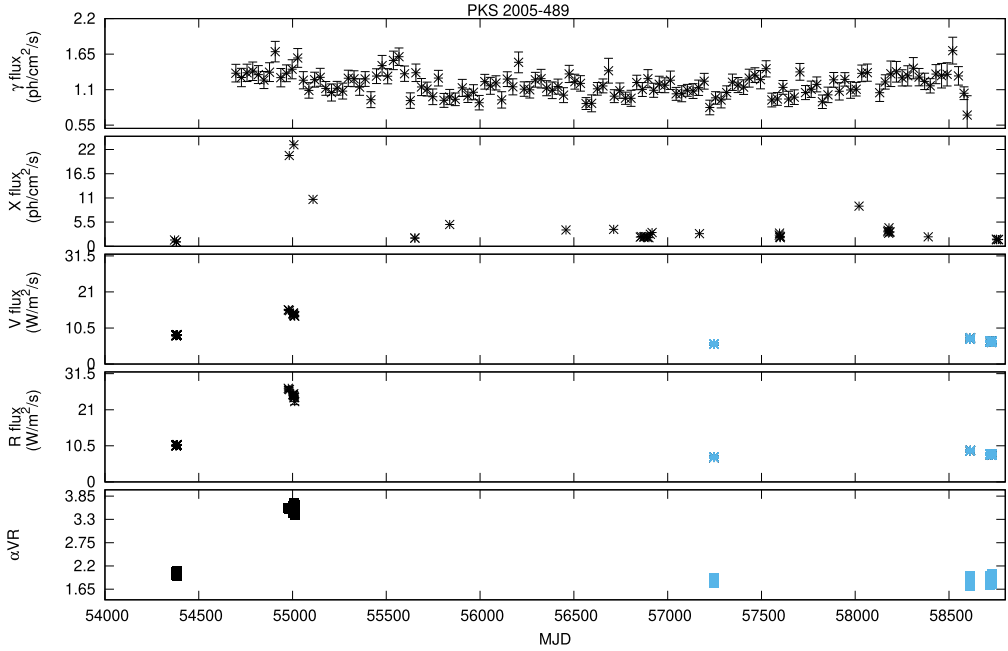

**Figure 14.** Multiwavelength light curves for the blazar PKS 2005−489 and its index $\alpha_{VR}$. The X-ray flux is in units of $10^{-11}$ ph cm$^{-2}$ s$^{-1}$. The $\gamma$-ray flux is in units of $10^{-7}$ ph cm$^{-2}$ s$^{-1}$.

physical settings of the jet, such as the injection of new relativistic particles or changes in the magnetic field strength, can affect both the synchrotron and inverse Compton emission. As a result, fluctuations in the synchrotron radiation at optical wavelengths can be associated with corresponding changes in the inverse Compton emission at $\gamma$-ray energies, leading to correlated variability between these bands. While the processes governing the emission at these two peaks may have different dependencies and time-scales, their interplay within the jet can produce correlated variability, as observed in the $\gamma$-ray and optical bands of some blazars.

To further confirm the correlation between our $\gamma$-ray and optical light curves, we proceed to repeat a similar experiment to that performed by Yuan et al. (2023). We merge both the $V$ and $R$ optical light curves into one, and we run a ZDCF algorithm (PYZDCF, Jankov et al. 2022) which estimates the error of each point by running several Monte Carlo simulations (10000 in our case). Both the optical and $\gamma$-ray light curves were normalized to their respective absolute maxima. We limited the space of possible time lags to $-365 \rightarrow 365$ d. In Fig. 12, we show the result of our ZDCF test. It differs from the one performed by Yuan et al. (2023) in the $\gamma$-ray light curve time bins, since we used the light curve from the 2FAVA catalogue, while they used their own Fermi data, and in the choice of software implementation. Our results are compatible to theirs within errors, with a maximum ZDCF value of $\sim 14^{+31}_{-6}$ d. This is consistent with the leptonic scenario in which the jet, which is intrinsically redder than the disc component, is dominating the spectrum; by being fed seed photons that increase the inverse Compton scattering (Ghisellini 2013; Sarkar et al. 2019b; Negi et al. 2022). Thus, optical and $\gamma$-ray emission are correlated through the emission mechanism, while X-rays are low due to the FSRQ nature of the source. In general, the more luminous blazars present their peaks towards lower energies, and can be modelled as having strong accretion feeding the jet (see Ghisellini et al. 2014; Arsioli & Chang 2018; Marchesini et al. 2019, and references therein).

The case of the BL Lac source PKS 2005−489 is less straightforward. Although it is known for having shown extreme X-ray flares (see e.g. Kapanadze 2021; Chase et al. 2023, and references therein), it has been in a quiet state for the last decade (Chase et al. 2023; Pininti et al. 2023). There was evidence that the X-ray flares may have been related to the optical variability state (Dominici et al. 2004), but since it entered the present quiet phase this could not be confirmed. Krauß et al. (2016) showed, however, that a peculiar deviation from the typical parabola shape is present in the X-ray band of the SED of this source. In our data, PKS 2005-489 exhibits significant X-ray variability during a period in which the γ-ray flux is steady. However, our optical data do not cover this period, not even when including the optical data published by the ATOM Data Base. The available optical data show very weak microvariability during only the first out of the three observation runs in 2015 August, and very weak IDV considering all the 2019 observations (see Table 3 and Fig. 3). We do detect strong variability in the time-scale of years, including both the 2015 and 2019 data sets, meaning that this source is undergoing a quiet phase in which short scale variability is mild or non-detected but noticeable variability can be detected over long periods of time. When this source showed variability in the optical band, it did so with a moderate BWB tendency ($r = 0.64$), in agreement with Ikejiri et al. (2011).

The *Fermi* γ-ray light curve is instead slightly decreasing, with an ∼ 38 per cent decrease in flux right before our optical data was taken. Our optical data also shows a slight decrease in flux, of ∼ 25 per cent. This is indeed consistent with the results published by Chase et al. (2023): The jet in PKS 2005−489 is undergoing a quiescent phase. The X-ray flare in our data is the 2009 flare, which they have attributed to either an external synchrotron emission component by secondary pairs generated by hadronic cascades in the broad-line region, so not to feed the inverse Compton, or to a strong magnetic field.

Being a BL Lac, PKS 2005−489 most probably has a thin accretion disc (see e.g. Paggi et al. 2009, and references therein), which is enough to power its jet in a classical SSC scenario. As an HSP source, the synchrotron component of PKS 2005−489 reaches its maximum between the UV and the optical wavelengths, depending on the epoch (Chase et al. 2023). Thus, its contribution is lower towards the red end of the spectrum. Indeed, Otero-Santos et al. (2022) showed that the optical spectrum of a variable BL Lac source can be modelled by two to four power laws, of which the main one is flat and the rest are bluer and steep. These components can increase or decrease, in proportion to the overall flux, following shocks generated within the jet (Li et al. 2015; Feng et al. 2020). It is expected that the disc is intrinsically blue. If the jet increases its emission, it would also turn bluer the overall spectrum, given the SED of this source.

## 6 CONCLUSIONS

We performed optical monitoring of 18 southern blazars, in the *V* and *R* filters, in order to analyse their variability state and to explore any relationship between flux variations and colour behaviour. The main results are summarized as follows:

(i) From the whole sample, 27.7 per cent of the blazars showed microvariability in the optical *V* band, while for the *R* band this percentage decreased to 22.2 per cent. Considering internight variability, 27.3 per cent of the blazars presented significant variations in both *V* and *R* filters. We also observed 5 of the 18 blazars in a time-scale of months, finding variability in four of them in both filters. And for two of the blazars in the sample, we observed significant variations in the *V* and *R* optical bands in time-scales of years.

(ii) For the sources where microvariability as well as variations in time-scales of days and months were detected, the analysis of the magnitude–magnitude diagrams together with the DCF showed that these variations in both *V* and *R* bands are produced by the photons emitted in the same region and by the same physical process.

(iii) We studied colour–magnitude correlations. We found that all six blazars that underwent flux variations in any of all the time-scales considered, showed a BWB trend at microvariability time-scales. Considering interday time-scales (days, weeks and months), a 50−50 per cent behaviour is obtained for both kind of trends (BWB–RWB). Finally, when a year time-scale is considered, only two objects remained: PKS 1510-089 (FSRQ), with an RWB trend, and PKS 2005-489 (BL Lac), with a BWB behaviour.

(iv) From the multiwavelength analysis, we conclude that PKS 1510−089, an FSRQ object, is undergoing an activity phase, which most likely is due to the jet being fed and dominating the flux spectrum. This can explain both its γ-ray and optical behaviour, as well as its RWB tendency. Our analysis shows that the γ-ray and optical light curves are cross-correlated with a lag compatible with less than 20 d, which is in agreement with previous results. On the contrary, PKS 2005−489, a BL Lac object, is in a quiescent state, in which it has been for more than a decade, in accordance to the literature. Its BWB moderate tendency could be explained due to its HSP nature, and/or to the presence of shocks within the jet.

## ACKNOWLEDGEMENTS

The present work was supported by grant 11/G178 from the Universidad Nacional de La Plata. This research has made use of the NASA/IPAC Extragalactic Database (NED) which is operated by the Jet Propulsion Laboratory, California Institute of Technology, under contract with the National Aeronautics and Space Administration. This paper has made use of up-to-date SMARTS optical/near-infrared light curves that are available at www.astro.yale.edu/smarts/glast/home.php. EJM would like to thank Dr R. I. Páez for the useful discussions on this work. EJM would like to thank the Summer School for Astrostatistics in Crete for providing training on the statistical methods adopted in this work. JAC acknowledge support by PIP 0113 (CONICET). JAC is CONICET researcher. JAC was also supported by grant PID2022-136828NB-C42 funded by the Spanish MCIN/AEI/ 10.13039/501100011033 and 'ERDF A way of making Europe' and by Consejería de Economía, Innovación, Ciencia y Empleo of Junta de Andalucía as research group FQM-322. We want to thank the anonymous referee for the useful and positive suggestions to improve this paper.

## DATA AVAILABILITY

The data underlying this article will be shared on reasonable request to the corresponding author.

This paper has been typeset from a TeX/LaTeX file prepared by the author.